\documentclass[a4paper,12pt]{article}

\usepackage[a4paper,
  left=2.5cm, right=2.5cm,
  top= 3cm, bottom=4cm]{geometry}
\usepackage{oldgerm}
\usepackage[dvips]{graphicx}
\usepackage{cite}
\usepackage{epic}
\usepackage{multirow}
\usepackage{calc}
\usepackage{amsmath,amssymb,amsthm,amscd}
\numberwithin{equation}{section}
\usepackage[bf]{caption}
\usepackage{longtable}
\usepackage{array}
\usepackage{enumerate}
\usepackage{float}
\usepackage{subfig}
\usepackage{mathtools}
\usepackage{multicol}
\usepackage{multirow}
\usepackage{epsfig}
\usepackage{graphicx}
\usepackage{pict2e}
\usepackage{color}
\usepackage{authblk}
\usepackage{cite}
\usepackage[all]{xy}
\usepackage[width=1.09\textwidth]{caption}
\usepackage{bbm}
\usepackage{hyperref}
\usepackage[makeroom]{cancel}

\usepackage[utf8]{inputenc}
\usepackage[T1]{fontenc}
\usepackage[english]{babel}
\usepackage{xspace}
\usepackage{pifont}
\usepackage{ytableau}
\usepackage{tikz}

\newcommand{\be}{\begin{equation}}  
\newcommand{\ee}{\end{equation}}

\newcommand{\rem}[1]{} 

\def\Z{\mathbb{Z}}

\def\Hirz[#1]{\mathbbm{F}_{#1}}
\def\o[#1]{\overline{#1}}

\def\cS{\mathcal{S}}

\def\L{\mathcal{L}}

\def\f{\textswab{f}}
\def\I{\textfrak{I}}

\newcommand{\U}[1]{\text{U(#1)}\xspace}
\newcommand{\SU}[1]{\text{SU(#1)}\xspace}

\newcommand{\beq}{\begin{equation}}
\newcommand{\eeq}{\end{equation}}
\newcommand{\bea}{\begin{eqnarray}}
\newcommand{\eea}{\end{eqnarray}}
\definecolor{verde}{rgb}{0.0, 0.5, 0.0}

\begin{document}

\begin{titlepage}
 

\vskip 2cm
\begin{center}
 
{\huge \bf \boldmath High U(1) charges in type IIB  models\\ \vspace{2mm} and their F-theory lift} 
 
 \vskip 2cm
\bigskip {
{{\bf  Francesco Mattia Cianci}$^{\,\text{\it a}}$,  
}
{{\bf  Dami\'an K. Mayorga Pe\~na}$^{\,\text{\it b}}$},\\
{{\bf Roberto Valandro}$^{\,\text{\it a},\,\text{\it c},\,\text{\it d}}$}
\bigskip }\\[3pt]
\vspace{0.cm}
{\it \small 
${}^{\text{a}}$ Dipartimento di Fisica, Universit\`a degli Studi di Trieste,\\ Strada Costiera 11, 34151 Trieste, Italy \\[8pt]
${}^{\text{b}}$Departamento de F\'{\i}sica, DCI, 
Universidad de Guanajuato,\\ Loma del Bosque 103, CP 37150,  Le\'on, Guanajuato, M\'exico \\[8pt]
${}^{\text{c}}$The Abdus Salam International Centre for Theoretical Physics,\\ Strada Costiera 11, 34151, Trieste,
Italy\\[8pt]
${}^{\text{d}}$ INFN, Sezione di Trieste, Via Valerio 2, 34127 Trieste, Italy\\[0.5cm]}

{\scriptsize \tt  francesco.cianci.phys[at]gmail.com, dmayorga[at]fisica.ugto.mx, roberto.valandro[at]ts.infn.it}
\\[2.4cm]
\end{center}

\vspace{3mm}

\begin{center} {\bf Abstract } \end{center}
\vspace{0.4Em}

We construct models with U(1) gauge group and matter with charges up to 6, in the context of type IIB compactifications. We show explicitly that models with charges up to 4 can be derived from corresponding models in F-theory by applying the Sen weak coupling limit. We derive which type IIB models should be the limit of charge 5 and 6 F-theory models. Explicit six dimensional type IIB  models with maximal charge 5 and 6 are constructed on an algebraic $K3$ surface that is the double cover of $\mathbb{CP}^2$. By using type IIB results we are also able to rediscover the F-theory charge 4 model in a straightforward way.

\end{titlepage}

\tableofcontents

\section{Introduction}

F-theory \cite{Vafa:1996xn,Morrison:1996na,Morrison:1996pp} constitutes a convenient framework to oversee the String Theory Landscape in various dimensions. In fact, provided the known dualities relating F-theory to type IIB string theory, heterotic string theory as well as M-theory,  F-theory compactifications are representative of  string theory vacua. In particular, F-theory is expected to be a good arena to study how effective field theories that look consistent at low energy may get obstructions when completed with quantum gravity. 


There has been more than a decade long search for consistent F-theory models that could be of relevance for particle physics, starting by the pioneering work of \cite{Beasley:2008dc,Donagi:2008ca}. This has led to 
significant progress towards understanding formal aspects of the theory related to the possible gauge symmetries (Abelian and non-Abelian, continuous and discrete), the type of matter that is allowed as well as the interaction terms. Regarding the moduli sector and the cosmological applications the theory still faces critical challenges. 
For this reason, the Landscape analysis in F-theory has concentrated essentially on issues related to  particle physics. 

A challenge for F-theory is to construct explicit models that are beyond the known landscape and, in doing so, identifying possible obstructions to realize lower dimensional effective theories. Regarding the non-Abelian symmetries, F-theory has an advantage over type IIB string theory in that it allows to obtain exceptional groups. Many of the constructed models  contain subgroups of $E_{8}$, but systematic surveys show that gauge symmetries including as many as $\mathcal{O}(100)$ $E_{8}$ factors are possible in four dimensions \cite{Halverson:2017ffz,Taylor:2017yqr}. Concerning the Abelian gauge symmetry sector in F-theory, significant progress in understanding the corresponding proper global setup  has led to fully fledged models with up to three $\U1$ factors \cite{Grimm:2010ez,Cvetic:2012xn,Mayrhofer:2012zy,Braun:2014qka,Cvetic:2013nia,Cvetic:2013qsa}. Similarly, it has been possible to obtain the Weierstrass form for globally consistent models with $\mathbb{Z}_2$, $\mathbb{Z}_3$ and $\mathbb{Z}_4$ discrete gauge symmetries \cite{Anderson:2014yva,Klevers:2014bqa,Mayrhofer:2014haa,Braun:2014qka,Lin:2015qsa,Cvetic:2016ner}.
   
A pressing question deals with what types of massless multiplets are possible in the effective theories resulting from F-theory. In dimesions higher than six, supersymmetry is constraining enough to ensure that the matter multiplet representations must not be bigger than the adjoint. In 6D and 4D one is only at the mercy of the anomaly cancellation conditions and these do not exclude the possibility for light exotic matter beyond the adjoint representation \cite{Kumar:2009us,Kumar:2009ae,Kumar:2009ac,Kumar:2010ru,Kumar:2010am,Seiberg:2011dr,Monnier:2017oqd,Taylor:2018khc}. Common representations that arise in F-theory models are the fundamental, the two index antisymmetric and the adjoint. However, in recent constructions it has been possible to obtain a three index symmetric and antisymmetric representations of various $\SU{N}$ gauge groups \cite{Grassi:2011hq,Anderson:2015cqy,Klevers:2017aku,Cvetic:2018xaq}. The problem becomes more severe when considering Abelian gauge symmetries: is there an upper bound on the maximum $\U1$ charge of a massless state? This question was raised in \cite{Park:2011wv,Taylor:2018khc,Raghuram:2018hjn}. So far, in F-theory it has been possible to construct globally consistent models with fields with $\U1$ charges only up to $Q=3$ and $Q=4$ \cite{Klevers:2014bqa,Raghuram:2017qut}. 

The above considerations are essentially an invitation to the explicit construction of F-theory models that lie at the frontier between the Landscape and the Swampland. In this paper we are going to approach the problem of \emph{constructing models with $\U1$ symmetries exhibiting massless states with large $\U1$ charges}.  Since $\U1$ symmetries are due to the presence of extra sections of the elliptic fibration, one expects the Weierstrass form describing these models to be highly specialized. $\U1$ models with charges $|Q|>2$ are described in terms of so called non unique factorization domains (non-UFD) \cite{Klevers:2017aku}. The non-UFD description has been used to extend the charge $Q=3$ model found in \cite{Klevers:2014bqa}, and to obtain a Weierstrass model with $Q=4$~\cite{Raghuram:2017qut}. 
We will show that these models are relatively easy to construct in the dual type IIB theory, in the perturbative limit. Thanks to this correspondence we will be able to straightforwardly extend the construction to charge $Q=5$ and $Q=6$ models in type IIB. Their F-theory lift is non-trivial, but in principle doable and we will leave this for a future project.

To understand the type IIB duals of F-theory models with one massless \U1 and states with high charge, we apply the Sen weak coupling limit \cite{Sen:1997gv}. This approach was intiated in \cite{MayorgaPena:2017eda}, where we obtained the weak coupling limit for a variety of globally consistent F-theory models, including one with gauge group $\U1$ and charge up to $Q=3$ and the $\mathbb{Z}_3$ models of \cite{Klevers:2014bqa}. In this work we proceed more systematically towards the analysis of the features of these models, including the $\U1$ construction with charge $Q=4$. Part of the systematics has to do with a way to construct divisors in the base $B$ of the elliptic fibration that are guaranteed to split into two divisors in the type IIB double cover Calabi-Yau (CY) $X$. Thanks to these techniques, we are able to obtain a family of type IIB models with a single massless $\U1$ symmetry and maximum charges that can reach up to $Q=6$. 
Thanks to type IIB we are also able to rediscover the F-theory charge $Q=4$ model of \cite{Raghuram:2017qut} in a straightforward way: in type IIB it is easy to understand which deformation of the D7-brane loci leads from a model with $\mathbb{Z}_3$ symmetry to the charge $Q=4$ model; applying the same deformation to the $\mathbb{Z}_3$ F-theory model of \cite{Klevers:2014bqa}, one immediately obtains the charge four model of \cite{Raghuram:2017qut}.

This paper is organized as follows, in Section~\ref{Sec:DoubCov} we describe the generalities of the double cover Calabi-Yau manifolds and the techniques necessary to construct D7-brane configurations with one massless \U1 and high charge spectrum. In Section~\ref{Sec:IIBFth} we discuss the features of type IIB models with higher $\U1$ charges, and present the configurations of branes and orientifolds leading to a single massless $\U1$ symmetry. We focus on the case where we have two \U1 D7-branes and one orientifold odd axion available to make one $\U1$ massive, leaving the other \U1 massless. We show that, under certain assumptions, in this setup the homology relations among the various 7-brane cycles allow models with maximum charges $Q=3,4,5,6$. In section~\ref{Sec:charge34} we describe the models with maximal charges $Q=3$ and $Q=4$ from the perspective of type IIB and F-theory. In Section~\ref{Sec:charge56} we present the type IIB versions for the charge $Q=5$ and $Q=6$. For the case of charge $6$ model we present an explicit $K3$ compactification. We devote Section~\ref{Sec:conclusion} to present our conclusions and prospects. We present some complementary material in the appendices.
   
\section{Type IIB manifold vs F-theory base}\label{Sec:DoubCov}

F-theory is defined on a manifold $Y$ that is an elliptic fibration over a base manifold $B$. The perturbative type IIB limit is defined on a double cover $X$ of the base manifold $B$. In this section, we will describe  tools that allow to directly connect the two descriptions.

\subsection{Double covers}

Our starting point will be a K\"ahler manifold $B$ and a line bundle $\L^{\otimes 2}$ on $B$. Take a section $b_2$ of $\L^{\otimes 2}$. The double cover $X$ of $B$ branched over the locus $b_2=0$ is given by the following hypersurface in the total space of the line bundle $\L$ \cite{Scwarzenberger,Friedman}:
\begin{equation}\label{Xeq}
X\,:\qquad \xi^2 = b_2 \:,
\end{equation}
where $\xi$ is a coordinate along the fiber. If the line bundle $\L$ is the anticanonical bundle of $B$, i.e. $\L=\bar{K}_B$, then $X$ is a Calabi-Yau space.

By construction, $X$ is symmetric under the involution 
\begin{equation}\label{Invol}
\textfrak{I} \,:\qquad \xi \mapsto - \xi \:,
\end{equation}
whose fixed point locus is $\{\xi=0\}$ (i.e. $b_2=0$). Taking the quotient of $X$ by the involution~\eqref{Invol} one obtains the space $B$ we started with.
Said in another way, there is a map 
\begin{equation}
 \f \,: \qquad X \,\rightarrow\, B \:,
\end{equation}
that is two-to-one, except at the fixed point locus.

Let us call Div$X$ the group of divisor on $X$ and Div$B$ the group of divisors on $B$. The map~$\textswab{f}$ induces a map 
\begin{equation}\label{fmapDivXB}
\f_\ast\,:\qquad \mbox{Div}X \,\rightarrow\, \mbox{Div}B\:.
\end{equation}
$\f_\ast$ is defined as follows on a reducible connected divisor $D$ of $X$: if $D$ is invariant under \eqref{Invol} then it is of degree two and $\f_\ast(D)=2\f(D)$; on the other hand, if $D$ is different from its image under \eqref{Invol}, then it is of degree one and $\f_\ast(D)=\f(D)$.\footnote{Notice that the map $\f_\ast \f^\ast$ acts as the multiplication by 2 on Div$B$, while $\f^\ast\f_\ast(D) = D + \textfrak{I}(D)$ for any divisor $D$ of $X$ (moreover, $D_X \cdot \f^\ast(D_B)=\f_\ast(D_X) \cdot D_B$ for $D_X\in$Div$X$ and $D_B\in\,$Div$B$).}

The map \eqref{fmapDivXB} induces a map between the Poincar\'e dual cohomologies: $\f_\ast :$ $H^2(X)\rightarrow H^2(B)$. If $b_2$ in \eqref{Xeq} is a sufficiently generic polynomial, then this map is one-to-one, i.e. $b^2(X)=b^2(B)$. If this is the case, then any pair of divisors $D$ and $\textfrak{I}(D)$ are in the same homology class. 
On the other hand, if $b_2$ presents a specific factorization, there can be more divisors in Div$X$ than in Div$B$.\footnote{This happens in two cases: 1) when the most generic section of $\L^{\otimes 2}$ must be factorized or 2) when one restricts the choice of the section $b_2$ to a factorized one. In the 
second case, if the manifold $X$ has dim$_{\mathbb{C}}>2$, then it is typically singular.}
In the last case, there will be a divisor $D_B$ of $B$ such that $\f^\ast (D_B)$ splits into two divisors $D_X$ and $\I(D_X)$ that are in different homology classes in $X$. Correspondingly, 
$b_2$ becomes a square when restricted to the locus $D_B$. If this happens globally, one can write $b_2$ as 
\begin{equation}\label{heqZi}
 b_2 \equiv \frac{s_6^2}{4} - s_Ls_R\:,
\end{equation}
where $s_6$ is a section of the line bundle $\L$, $s_L$ is a section of another line bundle $\L_L$ on $B$ and $s_R$ is consequently a section of $\L^{\otimes 2}\otimes \L_L^{-1}$. This factorization produces a  singularity on $X$ at $\xi=s_6=s_L=s_R=0$. This is a conifold singularity when $X$ is a three-fold.
However, in this paper we will consider type IIB compactifications to 6 dimensions; hence $X$ is a (complex) two-dimensional space in a three-dimensional ambient space and then there are generically no solution to the four equations. For this reason, in this paper we will assume $X$ to be smooth. Most of our results will be valid also for 4D compactifications, if one considers base manifolds $B$ such that the locus $\{s_6=s_L=s_R=0\}$ is empty  \cite{Krause:2012yh}.

Let us consider the divisor $D_{B}=\{s_L=0\}$ in $B$. $\f^\ast (D_B)$ splits on $X$ into the loci $\{\xi-s_6/2=s_L=0\}$ and $\{\xi+s_6/2=s_L=0\}$, that are in different homology classes \cite{Krause:2012yh,Collinucci:2012as,Braun:2014nva}. Let us call the corresponding divisors $D_{L,-}$ and $D_{L,+}$. They are image to each other under the involution $\xi\mapsto - \xi$ and by construction their homology classes satisfy the following relation in $H^2(X)$
\begin{equation}\label{DLpmsL}
 [D_{L,+}] +  [D_{L,-}] = [s_L]  \:,
\end{equation}
where $[s_L]$ is the homology class of the locus $\{s_L=0\}$ in $X$. 

In the following, we will need information about the intersection numbers. As before, we take $X$ to be a (complex) two-dimensional hypersurface in the ambient space $A$ that is the total space of the line bundle $\L$ on $B$. If $P$ is a polynomial in the coordinates of $B$, then we have 
\begin{equation}\label{PDLp}
[P]\cdot [D_{L,\pm}] = [P]\cdot [s_L]\cdot [\xi] |_A = \frac12 [P]\cdot[s_L] 
\end{equation}
where the double intersections are always  meant in $X$ if not specified. We keep this convention also in the following.
Moreover, we have
\begin{eqnarray}\label{DLpDLm}
 [D_{L,+}]\cdot [D_{L,-}]   &=& [s_L]\cdot [\xi] \cdot [s_6] |_A = \frac12 [s_L] \cdot c_1(\L) = [D_{L,\pm}]\cdot [\xi]
\end{eqnarray}
and, using \eqref{DLpmsL},
\begin{eqnarray}\label{DLpSquare}
 [D_{L,\pm}]^2  &=& ([s_L]-[D_{L,\mp}])\cdot [D_{L,\pm}]  = \frac12 [s_L] \cdot ([s_L] - c_1(\L) )  \:.
\end{eqnarray}

\subsection{Splitting divisors from matrices}\label{Sec:splittingDiv}

The existence of a new divisor class can be detected also from the fact that if $b_2$ takes the form \eqref{heqZi}, then  $X$ has a $2\times 2$ matrix factorization (MF)(see \cite{Collinucci:2016hgh} for more details on an analogous model),\footnote{For a nice review on MF in physics see \cite{Jockers:2007ng}. For application of MF in similar contexts see also \cite{Braun:2011zm,Collinucci:2014taa}.} i.e. there exist matrices $(\mathrm{M},\tilde{\mathrm{M}})$ such that
\begin{equation}
 \mathrm{M}\cdot \tilde{\mathrm{M}} = \tilde{\mathrm{M}}\cdot \mathrm{M} = (-\xi^2+b_2) \mathbbm{1}_2 \:,
\end{equation}
with 
\begin{equation}\label{MF2X}
\mathrm{M} = \left( \begin{array}{cc}
 s_R  & -\xi-\frac{s_6}{2}\\   \xi-\frac{s_6}{2} & s_L \\
\end{array}\right) \qquad \mbox{and} \qquad 
\tilde{\mathrm{M}} = \left( \begin{array}{cc}
s_L & \xi+\frac{s_6}{2} \\ -\xi+\frac{s_6}{2} & s_R \\ 
\end{array}\right)   \:.
\end{equation}
One can then define
\begin{eqnarray}
\label{Mseq}&& \mathcal{L}_{\mathrm{M}} \equiv \mbox{coker} \left( \, V_2 \,\,\stackrel{\mathrm{M}}{\longrightarrow}\,\,  W_2 \,\right) \:, 
\end{eqnarray}
where $V_2=L\otimes ( \L_L^{-1} \oplus \L^{-1} )$ and $W_2=L\otimes \L_L^{-1}\otimes ( \L \oplus \L_L )$ (with $L$ an arbitrary line bundle). All the involved line bundles are naturally defined on the base manifold $B$ and can be easily lifted to $X$  by the map $\f^\ast$.

If $X$ is smooth, $\L_{\mathrm{M}}$ is a line bundle on $X$: in fact det$\mathrm{M}=\xi^2-b_2$ and hence at~$\xi^2=b_2$ the matrix rank goes down to one, generating a one-dimensional cokernel.\footnote{If the locus $\{\xi=s_6=s_L=s_R=0\}$ is non-empty, then on this points the matrix rank goes down by two units and at that point the cokernel dimension jumps from one to two. Such a $\L_{\mathrm{M}}$ is called a non-trivial irreducible Maximal Cohen Macaulay (MCM) module.}
The first Chern class of a line bundle $\mathcal{L}$ is the Poincar\'e dual of the divisor where a generic section of $\mathcal{L}$ vanishes. The line bundle $\mathcal{L}_{\mathrm{M}}$ is given as the cokernel of the map $\mathrm{M}$.
For a $2\times 2$ MF $(\mathrm{M},\tilde{\mathrm{M}})$, one has an isomorphism  coker$\mathrm{M}\cong$im$\tilde{\mathrm{M}}$ given by the map $\tilde{\mathrm{M}}$ restricted to coker$\mathrm{M}$.\footnote{The space coker$\mathrm{M}$ is given by $W_2/$im$\mathrm{M}$. But over $\xi^2=b_2$ the exactness of the sequence implies that  im$\mathrm{M}=$ker$\tilde{\mathrm{M}}$. Hence, by definition, ker$\tilde{\mathrm{M}}=0$ when the map $\tilde{\mathrm{M}}$ is restricted to coker$\mathrm{M}$ and hence it is an isomorphism between coker$\mathrm{M}$ and im$\tilde{\mathrm{M}}$.}
Hence the locus where a section of $\mathcal{L}_{\mathrm{M}}$ vanishes is the same as the locus where a section of im$\tilde{\rm M}$ vanishes, i.e.
\begin{eqnarray}\label{DivInDC}
&& c_1(\mathcal{L}_{\mathrm{M}})= [D_{p}] \qquad\mbox{with} \qquad D_p\,:\,\,\,\,\,\tilde{\mathrm{M}}\cdot \left( \begin{array}{c} p_1 \\ p_2 \\ \end{array}\right)=0  \:, 
\end{eqnarray}
where $\vec{p}=(p_1,p_2)$ is a section of the vector bundle $W_2$ in \eqref{Mseq}.  The homology class of $D_p$ can be computed in the following way: one can deform the generic divisor $D_p$ by setting $p_2=0$. Then the ideal $\tilde{\mathrm{M}} \vec{p}=0$ splits into the union of $p_1=0$ in $X$ and the divisor $D_{L,-}$, i.e. at the level of homology classes:
\begin{equation}\label{classDp}
  [D_p] = [p_1] + [D_{L,-}] \:.
\end{equation}

The equation $\xi^2-b_2=0$ has an inequivalent MF $(\mathrm{M}',\tilde{\mathrm{M}}')$ that is obtained from \eqref{MF2X} by taking $\xi\mapsto -\xi$.
Following the same steps as before, we can then construct a line bundle $\L_{M'}$, whose first Chern class is 
\begin{eqnarray}\label{DivInDCp}
&& c_1(\mathcal{L}_{\mathrm{M}'})= [D_{p}'] \qquad\mbox{with} \qquad D_p'\,:\,\,\,\,\,\tilde{\mathrm{M}}'\cdot \left( \begin{array}{c} p_1 \\ p_2 \\ \end{array}\right)=0  
\end{eqnarray}
whose homology class is 
\begin{equation}\label{classDpp}
  [D_p'] = [p_1] + [D_{L,+}]  \:.
\end{equation}
In particular, when $(p_1,p_2)=(1,0)$ we have $D_{p}=D_{L,-}$ and $D_{p'}=D_{L,+}$. 

Notice  that 
\begin{equation}\label{eq:Dp}
 \tilde{\mathrm{M}} = \mathrm{A}_1+\xi\,\mathrm{B}_0 \qquad \mbox{where} \qquad \mathrm{A}_1 = \left( \begin{array}{cc}
s_L & \frac{s_6}{2}  \\  \frac{s_6}{2} & s_R \\
\end{array}\right) \mbox{ and } \,\,\mathrm{B}_0= \begin{pmatrix}
0 & 1 \\ -1 & 0 \\
\end{pmatrix}\:.
\end{equation}
 $\tilde{\mathrm{M}}'$ is written also in terms of $\mathrm{A}_1$ and $\mathrm{B}_0$ but with $\xi\mapsto -\xi$.
It is then easy to see that the divisors $D_p$ and $D_p'$ intersect each other over two loci:
\begin{equation}\label{DD'AwayO7}
  \{ \xi = 0 , \,\,\mathrm{A}_1\cdot\vec{p}=0\} \qquad\mbox{ and }\qquad \{p_1=0, \,p_2=0\} \:.
\end{equation}
The first one is on top of the fixed point locus of the involution $\I$, while the other may intersect the fixed point locus but its points are generically not fixed.

The two divisors $D_p$ and $D_{p'}$ are mapped to each other by the involution $\I$. Hence they will be projected down by $\mathfrak{f}$ to the same divisor $D_{B,p}$ of $B$: $D_{B,p}=\f_\ast D_{p}=\f_\ast D_{p}'$, where we also have $\f^\ast D_{B,p}=D_p+D_p'$.

We now prove that $D_{B,p}$ is described by the equation\footnote{Notice that the submanifold \eqref{DBp1p2} in $B$ is singular at $p_1=p_2=0$. This is not surprising. The locus $\{p_1=p_2=0\}$ is the intersection locus of the divisor $D_p$ and $D_p'$ away from the fixed point locus. On the other hand, the two divisors of $X$ join each other in $B$, forming a connected divisor $D_{B,p}$. The two branches of $D_{B,p}$ still intersect transversally at $p_1=p_2=0$, hence generating a singularity.
This also happens when $D$ and $D'$ are in the same homology class.
}
\begin{equation}\label{DBp1p2}
 (p_1,p_2)\cdot \mathrm{A}_1\cdot \left( \begin{array}{c} p_1 \\ p_2 \\ \end{array}\right)=0 
 \:.
\end{equation}
\begin{itemize}
\item We first prove that all points of $D_{B,p}$ satisfy \eqref{DBp1p2}: such points are pulled-back by $\f^\ast$ to points either of $D_p$ or of $D_p'$; over these loci $\mathrm{A}_1\vec{p}=\mp \xi \mathrm{B}_0\vec{p}$. Hence  $\vec{p} 	\cdot  \mathrm{A}_1\vec{p}=\mp \xi  \vec{p} 	\cdot \mathrm{B}_0\vec{p}=0$ as $\mathrm{B}_0$ is antisymmetric.

\item We then prove that any points satisfying \eqref{DBp1p2} belong to $D_{B,p}$: over these points, $\mathrm{A}_1\vec{p}$ is orthogonal to $\vec{p}$, i.e. it is  proportional to $\mathrm{B}_0\,\vec{p}$. This means that $\mathrm{B}_0\,\vec{p}$ is an eigenvector of $\mathrm{A}_1\mathrm{B}_0$ ($\mathrm{B}_0^2=-\mathbbm{1}$). We also have $(\mathrm{A}_1\mathrm{B}_0)^2=-$det$(\mathrm{A}_1) \mathbbm{1}$. Hence the eigenvalues of $\mathrm{A}_1\mathrm{B}_0$ are $\pm \sqrt{b_2}$. If we pull-back these points, they will belong either to $D_p$ or to $D_p'$ (since on $X$ we have $b_2=\xi^2$).
\end{itemize}

The formula \eqref{DBp1p2} gives then an algebraic expression for a divisor $D_{B,p}$ of $B$ that, once lifted to the double cover $X$, splits into two components, one the image of the other under the inolution $\I$.\footnote{It is easy to see why the  equation \eqref{DBp1p2} splits when intersected with $\xi^2=b_2$. In fact, $b_2=$-det$\mathcal{A}$. Hence, on $X$ the determinant of $\mathcal{A}$ is a square and then the quadratic form in \eqref{DBp1p2} factorizes into two factors (that are exchanged by taking $\xi\mapsto -\xi$).}

\subsection{Splitting divisors of higher degree}

The procedure outlined above can be used to construct other pairs of algebraic cycles mapped to each other by the involution $\I$:
Take the line bundle $\mathcal{L}_{\mathrm{M}}^{\otimes 2}$; thanks to the isomorphism given by $\tilde{\mathrm{M}}$ restricted on coker$\mathrm{M}$, this line bundle is isomorphic to im$(\tilde{\mathrm{M}}\otimes \tilde{\mathrm{M}})$. The vanishing locus of a generic section is then given by 
\begin{equation}\label{vanishLocL2}
  \tilde{\mathrm{M}}\otimes \tilde{\mathrm{M}} \cdot \begin{pmatrix}
  q_1\\q_2\\q_3\\q_4\\
  \end{pmatrix}=0\:.
\end{equation}
On top of the double cover $X$, the matrix $\tilde{\mathrm{M}}\otimes \tilde{\mathrm{M}}$ has rank$\left(\tilde{\mathrm{M}}\otimes \tilde{\mathrm{M}}\right)=$rank$(\tilde{\mathrm{\mathrm{M}}})^2=1$. Hence the condition \eqref{vanishLocL2} is actually giving a codimension one locus, as it should be for a divisor. 

The $4\times 4$ matrix $\tilde{\mathrm{M}}\otimes \tilde{\mathrm{M}}$ can be effectively reduced to a $3\times 3$ matrix. This can be shown for a generic matrix $\mathcal{M}=\begin{pmatrix}
a & b \\ c & d \\
\end{pmatrix} $. The tensor produc of $\mathcal{M}$ with itself is
$$
\mathcal{M}\otimes \mathcal{M} = \begin{pmatrix}
a^2 & ab & ab & b^2 \\ ac & ad & bc & bd \\  ac & cb & ad & bd \\  c^2 & cd & cd & d^2 \\  
\end{pmatrix} \:,
$$
that is equivalent to the block diagonal form
$$
\begin{pmatrix}
a^2 & -ab & b^2 & \\ -ac & \frac12(ad+cb) & -bd & \\ c^2 & -cd & d^2 & \\ & & & 2(ad-cb)\\
\end{pmatrix}
$$
up to operation of summing or subtracting lines or columns and changing the order of lines and columns. Notice that the element $2(ad-bc)$ is proportional to the determinant of $\mathcal{M}$. Let us apply this to $\mathcal{M}=\tilde{\mathrm{M}}$.
Notice that in the block-diagonal form the relevant part of $\tilde{\mathrm{M}}\otimes \tilde{\mathrm{M}}$ is the $3\times 3$ block (since  det$\tilde{\mathrm{M}}$ vanishes on $X$). We can moreover separate the parts linear in $\xi$ as 
\begin{equation}
(\tilde{\mathrm{\mathrm{M}}}\otimes \tilde{\mathrm{\mathrm{M}}})_{3\times 3} =\mathrm{A}_2 + \xi \,\mathrm{B}_1
\end{equation}
with
\begin{equation}
\mathrm{A}_2=\begin{pmatrix}
s_L^2  &  -\frac{s_6s_L}{2}   &  \frac{s_6^2}{2}-s_Ls_R \\
-\frac{s_6s_L}{2}  &  s_Ls_R  &  -\frac{s_6s_R}{2}  \\
\frac{s_6^2}{2}-s_Ls_R  &  -\frac{s_6s_R}{2}  &  s_R^2 \\
\end{pmatrix}\qquad\mbox{and}\qquad
\mathrm{B}_1=\begin{pmatrix}
0  &  -s_L  &  s_6  \\  s_L  &  0  & - s_R  \\  - s_6  &  s_R  &  0  \\
\end{pmatrix} \:.
\end{equation}
The subscript ``$2$'' signals the fact that $\mathrm{A}_2$ is homogeneous of degree 2 in $s_L,s_6,s_R$. On the other hand $\mathrm{B}_1$ is of degree 1.

The two divisors mapped to each other by $\I$ are then
\begin{equation}
 D^{(2)}_q : \,\, (\mathrm{A}_2+\xi \,\mathrm{B}_1)\cdot  \begin{pmatrix}
  q_1\\q_2\\q_3\\
  \end{pmatrix}=0\qquad\mbox{and}\qquad
  D^{(2)'}_q : \,\, (\mathrm{A}_2-\xi \,\mathrm{B}_1)\cdot  \begin{pmatrix}
  q_1\\q_2\\q_3\\
  \end{pmatrix}=0 \:.
\end{equation}
Their divisor classes can be derived as above and are 
\begin{equation}
 [D_q^{(2)}]  = 2 [D_{L,-}]  + [q_1] \qquad\mbox{ and }\qquad
 [D_q^{(2)'}] = 2 [D_{L,+}]  + [q_1]\:.
\end{equation}
The two divisors $D^{(2)}_q$ and $D^{(2)'}_q$ intersect each other over  two loci:
\begin{equation}\label{order2D7D7pintersection}
\{ \xi=0, \mathrm{A}_2\cdot \vec{q}=0\} \qquad \mbox{ and } \qquad \{ \mathrm{B}_1\cdot \vec{q}=0\}\:.
\end{equation}
The points of the second locus are generically away from the fixed point locus. To write the second equation we have used the fact that 
\begin{equation}\label{ArispB}
\mathrm{A}_2= \mathcal{I}
\cdot \mathrm{B}_1 \quad {\rm where} \quad \mathcal{I}= \mathrm{B}_1^t \cdot C\,, \quad \mbox{and} \quad     {\scriptsize C = \begin{pmatrix}
0&0& -\frac{1}{2}\\ 0&1&0 \\  -\frac{1}{2}&0&0 \\
\end{pmatrix}} \:.
\end{equation}

One can then find the divisor of $B$ that splits into $D^{(2)}_q$ and $D^{(2)'}_q$ when lifted to the double cover $X$, i.e. 
$D_{B,q}^{(2)}=\f_\ast D_{q}^{(2)}=\f_\ast D_{q}^{(2)'}$, where we also have $\f^\ast D_{B,q}^{(2)}=D_{q}^{(2)}+D_{q}^{(2)'}$. Following  similar considerations as for the previous case, we found that it is given by
\begin{equation}\label{DB2inquot}
D_{B,q}^{(2)} \,:\,\,\, \begin{pmatrix}
 q_1 & q_2 & q_3 \\
\end{pmatrix} \cdot \mathrm{A}_2 \cdot   \begin{pmatrix}
 q_1 \\ q_2 \\ q_3 \\
\end{pmatrix} 
=  0 \:.
\end{equation}

By analogous considerations, one can construct divisors $D_{r}^{(3)}$ and $D_{r}^{(3)'}$:   the matrix $\tilde{\rm M}\otimes \tilde{\rm M}\otimes \tilde{\rm M}$ can be reduced to a $4\times 4$ matrix $A_3+\xi B_2$ on top of $X$, where
\begin{equation}\label{eq:m3LR}
A_3={\scriptsize\begin{pmatrix}
s_L^3 & -s_L^2 s_6/2 & (s_L s_6^2 - 2 s_L^2 s_R)/2 & (-s_6^3 + 3 s_L s_6 s_R)/2 \\
-s_L^2 s_6/2 & s_L^2 s_R & -s_L s_6 s_R/2 & (s_6^2 s_R - 2 s_3 s_R^2)/2 \\
(s_L s_6^2 - 2 s_L^2 s_R)/2 & -s_L s_6 s_R/2 & s_L s_R^2 & -s_6 s_R^2/2 \\
(-s_6^3 + 3 s_L s_6 s_R)/2 & (s_6^2 s_R - 2 s_L s_R^2)/2 & -s_6 s_R^2/2 & s_R^3 
\end{pmatrix}}
\end{equation} 
and\footnote{
Once again there is a relation between $A_3$ and $B_2$: 
$
A_3=
{\scriptsize\begin{pmatrix}
s_6/2& s_L& 0& 0\\
-s_R/2& 0& s_L/2& 0\\
0& -s_R/2 & 0& s_L/2 \\
 0& 0& -s_R& -s_6/2
\end{pmatrix}}\cdot B_2
$
}
\begin{equation}
B_2={\scriptsize\begin{pmatrix}
0 & -s_L^2 & s_6 s_L & -s_6^2 + s_L s_R \\
 s_L^2 & 0 & -s_L s_R & s_6 s_R \\
-s_6 s_L & s_L s_R & 0 & -s_R^2 \\
 s_6^2 - s_L s_R & -s_6 s_R & s_R^2 & 0 
\end{pmatrix}} \:.
\end{equation} 
The two divisors are then
\begin{equation}\label{eq:Dr}
 D^{(3)}_r : \,\, (\mathrm{A}_3+\xi \,\mathrm{B}_2)\cdot  \begin{pmatrix}
  r_1\\r_2\\r_3\\r_4\\
  \end{pmatrix}=0\qquad\mbox{and}\qquad
  D^{(3)'}_r : \,\, (\mathrm{A}_3+\xi \,\mathrm{B}_2)\cdot  \begin{pmatrix}
  r_1\\r_2\\r_3\\r_4\\
  \end{pmatrix}=0
\end{equation}
and their homology classes are
\begin{equation}
 [D_q^{(3)}]  = 3 [D_{L,-}]  + [r_1] \qquad\mbox{ and }\qquad
 [D_q^{(3)'}] = 3 [D_{L,+}]  + [r_1]\:.
\end{equation}
The divisor $D_{B,q}^{(3)}=\f_\ast D_{q}^{(3)}=\f_\ast D_{q}^{(3)'}$ is given by the equation
\begin{equation}\label{DB2inquot3}
D_{B,q}^{(3)} \,:\,\,\, \begin{pmatrix}
 r_1 & r_2 & r_3 & r_4 \\
\end{pmatrix} \cdot \mathrm{A}_3 \cdot   \begin{pmatrix}
 r_1 \\ r_2 \\ r_3 \\ r_4 \\
\end{pmatrix} =  0 \:.
\end{equation}

One can in principle continue with this procedure to obtain divisors in homology classes $n[D_{L,\mp}] + [P]$. We give the result for $n=4$ in Appendix~\ref{App:matrixesn4}.

\subsection{Odd divisor classes}
\label{sec:odd}

The involution $\I$ in \eqref{Invol} splits the second cohomology of $X$ into even and odd elements: $H^2(X)=H^2_+(X)\oplus H^2_-(X)$. The even elements, as we said, are in one-to-one correspondence with the divisors of $B$ ($b^2(B)=b^2_+(X)$).  The odd elements can be written as differences between divisors (or Poincar\'e dual two-forms) that are mapped to each other by $\I$. Of course, to obtain non-trivial elements, one needs that such divisors are in different classes. The divisors constructed in Section~\ref{Sec:splittingDiv} are of this type. 
We can then associate an odd class in $H^2(X)$ with the matrix factorization \eqref{MF2X}:
\begin{equation}
D_- \equiv [D_p]-[D_p']  \:.
\end{equation}
Changing $\vec{p}$ does not affect the class of $D_-$ in $H^2(X)$, as can be seen from \eqref{classDp} and~\eqref{classDpp}:
\begin{equation}
D_-=  [D_p]-[D_p'] = [D_{L,-}]-[D_{L,+}]=[D_{R,+}]-[D_{R,-}] \qquad\forall\vec{p}\:.
\end{equation}

If we take differences of connected algebraic divisors $D_q^{(n)}$ and $D_q^{(n)'}$, constructed by the matrices $\tilde{\mathrm{M}}^{\otimes n}$ and $(\tilde{\mathrm{M}}')^{\otimes n}$,
we obtain multiples of $D_-$:
$$
[D_{q}^{(n)}] - [D_{q}^{(n)'}] = n([D_{L,-}]-[D_{L,+}]) = n D_- \:.
$$

If the space $X$ admits further  MF's,  we can associate an independent odd class to each of them.

\section{Type IIB limit of F-theory}\label{Sec:IIBFth}

\subsection{F-theory models with \U1 gauge group}

F-theory is defined on a CY manifold $Y$ that is an elliptic fibration over a K\"ahler manifold $B$. If the fibration has a section (called the ``zero section'') the space $Y$ can be described by a Weierstrass model, i.e. by the equation
\begin{equation}\label{Weierstrass}
y^2 = x^3+fxz^4+gz^6,
\end{equation}
where $f$ and $g$ are sections of $\bar{K}_B^{\otimes 4}$ and $\bar{K}_B^{\otimes 6}$ respectively ($K_B$ is the canonical line bundle of the base manifold $B$), and $x$, $y$ and $z$ sections of $(\bar{K}_B\otimes H)^{\otimes 2}$, $(\bar{K}_B\otimes H)^{\otimes 3}$ and $H$  ($H$ is the line bundle  which $z$ belongs to). The elliptic curve degenerates over the zero locus of the discriminant $\Delta=4f^3+27g^2$: this gives the location of the 7-branes.

If the Weierstrass model is smooth, the effective lower dimensional theory has no gauge group nor matter. In this paper, we are interested on the simplest gauge group, that is \U1. This is realized when the elliptic fibration $Y$ has one extra section. When this happens, $Y$ develops singularities along codimension-2 loci in the base $B$, where states charged under the \U1 gauge group live. The charge of states localized at different loci are typically different. 

So far, models with charges up to $4$ have been constructed as global F-theory compactifications \cite{Grimm:2010ez,Morrison:2012ei,Klevers:2014bqa,Raghuram:2017qut}.
The Weierstrass model descriptions of such configurations are birationally equivalent to smooth manifolds $\tilde{Y}$ that are hypersurfaces in an ambient space $ \mathbb{P}^B_{F_i}$ that is the fibration of a toric two-dimensional variety $\mathbb{P}_{F_i}$ over the base manifold $B$. They are described by the equation \cite{Batyrev:1994hm,Klevers:2014bqa}, 
\begin{eqnarray}\label{batyrev}
	P(z_1,...,z_k)=\sum_{w\in F_i^\ast}s_w \prod_{\ell=1}^{k}z_{\ell}^{\langle v_{\ell},w\rangle+1}=0\:.
\end{eqnarray}
The toric variety $\mathbb{P}_{F_i}$ is defined over the polytope $F_i$, with homogeneous coordinates $z_1,...,z_k$, one per each non-zero lattice vector $v_\ell\in F_i$, and $w$ being lattice vectors in the dual polytope $F_i^*$ (including the origin).  The polynomial in \eqref{batyrev}, called Batyrev polynomial, defines a hypersurface $X_{F_i}\in \mathbb{P}^B_{F_i}$ elliptically fibered over the base $B$. The accompanying coefficients $s_w$ are taken as sections of line bundles over the base manifold $B$ and and can be seen (locally) as polynomials on the base manifold's coordinates.
The birational map allows to write $f$ and $g$ in \eqref{Weierstrass} in terms of the sections~$s_w$: the particular expression of $f$ and $g$ brings all the information about which configuration one has, i.e. one can deform the $s_w$, by choosing a different generic section in the same line bundle, but the gauge group and which charged spectrum one has, does not change. Instead, if some of $s_w$'s are identically zero or have very specific factorized forms, then the gauge group or the charged spectrum can change. 
This is what happens for example for the charge 4 models \cite{Raghuram:2017qut}: As we will see in Section~\ref{Sec:Charge4FromTth}, one can start from  a model with a $\Z_3$ discrete symmetry in the form \eqref{batyrev} and deform some of the $s_w$'s to specific sections of the corresponding line bundle; these $s_w$'s will be written in terms of sections $a_1,b_1,d_i$ of new line bundles on the base manifold $B$ generating a model with gauge group \U1 and charge 4 matter. Now choosing different generic sections $s_w',a_1,b_1,d_i$ in the  new line bundles does not chage the gauge group and matter spectrum. In the following  the sections defining the gauge group and the matter sector will be called $\mathsf{s}_\kappa$ (so, for example, in \cite{Raghuram:2017qut} $\mathsf{s}_\kappa=s_w',a_1,b_1,d_i$).


\subsection{Sen limit}\label{Sec:SenLim}

In this paper we are interested in the weak coupling limit of F-theory comapactifications with \U1 gauge group. This limit, first studied by Sen \cite{Sen:1997gv}, is a limit in the complex structure moduli space. For this reason, it is a delicate limit: complex structure deformations can change the gauge group and the matter spectrum of the F-theory model; on the other hand, for a weak coupling limit one means studying the F-theory 7-brane configuration under consideration but  in  the perturbative type IIB language. Hence, the weak coupling limit should not change the gauge group and the matter spectrum. As we have said above, the information about the 7-brane configuration is encoded in 
a choice of line bundles over $B$ and corresponding generic sections $\mathsf{s}_\kappa$, in terms of which $f$ and $g$ are expressed. So, the Sen limit should not deform the polynomials $\mathsf{s}_\kappa$.

In order to see how the Sen limit works, one can first reparameterize
 $f$ and $g$ in \eqref{Weierstrass} as
	\begin{eqnarray}
	\label{def_b}
	f=-\frac{b_2^2}{3}+2b_4, \quad g=\frac{2}{27}b_2^3-\frac{2}{3}b_2b_4+b_6,
	\end{eqnarray}
where $b_i$'s are sections of $\bar{K}_B^{\otimes i}$. Correspondingly the discriminant $\Delta=4f^3+27g^2$ becomes
	\begin{eqnarray}
	\Delta=4b_2^2(b_2b_6-b_4^2)-36b_2b_4b_6+32b_4^3+27b_6^2.
	\end{eqnarray}
	
In F-theory, the type IIB axio-dialaton $\tau$ (and thus the string coupling) varies over the base manifold $B$. The $SL(2,\mathbb{Z})$ invariant function 	$j(\tau)$ is in fact given by $f^3/\Delta$.
Sen found a limit that sets the string coupling small almost everywhere in $B$: if one scales 
	\begin{eqnarray}
	\label{scaling_b}
	b_2\rightarrow \epsilon^0 b_2, \quad b_4\rightarrow \epsilon^1 b_4, \quad b_6 \rightarrow \epsilon^2 b_6,
	\end{eqnarray}
the discriminant becomes
\begin{eqnarray}\label{discrWC}
	\Delta& \xrightarrow[\epsilon\rightarrow 0]{}& -4\epsilon^2 b_2^2 (b_4^2-b_2b_6)+\mathcal{O}(\epsilon^3),
\end{eqnarray}
and then the string coupling becomes small: In fact $j(\tau)\xrightarrow[]{\epsilon\rightarrow 0} \epsilon^{-2}$ and $j(\tau)= \exp{(-2\pi i \tau)}+744+\mathcal{O}[\exp{(2\pi i \tau)}]$ (recall that $\tau=C_0+i \,e^{-\phi}$ with $g_s=e^\phi$).

As it can be seen from \eqref{discrWC},  in this limit the codimension-1 loci of the base where the 7-branes lie are described by the two zeroes of the discriminant:
	\begin{eqnarray}
	b_2=0\,\,\, \quad\mbox{ and }\qquad\,\,\, \Delta_E\equiv b_4^2-b_2b_6=0.
	\end{eqnarray}
From the monodromies of  $\tau$ around these  loci, one can find that  $b_2=0$ and $\Delta_E=0$ describe respectively an O7-plane and a D7-brane, i.e. one has only perturbative objects.

The type IIB compactification manifold must be a double cover of the base $B$ branched over the O7-plane locus, i.e. it is given by the equation $\xi^2=b_2$ in \eqref{Xeq}. The involution $\I$ that sends $\xi\mapsto -\xi$ is the orientifold involution.

When the Weierstrass model is smooth, the locus $\Delta_E=0$ describes one brane that is invariant with respect to the involution $\xi\rightarrow -\xi$; if the $b_i$ have a proper special form, then $\Delta_E$ can factorize so that there is more than one stack of D7-branes. As we will see, one can also have pairs of branes and their orientifold images. 

Let us start with a model of the form \eqref{batyrev} (or specializations thereof). There is a birational map to the Weierstrass model that gives $f$ and $g$ as functions of $\mathsf{s}_\kappa$. After a choice of $b_2$, that will also be a function of $\mathsf{s}_\kappa$ (it is not a coincidence the name we gave to the polynomials in \eqref{heqZi}), one can derive the expressions for $b_4$ and $b_6$ in terms of $\mathsf{s}_\kappa$: using \eqref{def_b}, we need to take 
$b_4=(f+b_2^2/3)/2,$ and then $b_6=g-(2/27)b_2^3+(2/3) b_2 b_4.$
These will be functions of the $\mathsf{s}_\kappa$'s as well. As we said above, in the weak coupling limit we should not deform the polynomial $\mathsf{s}_\kappa$ as they bring the information about the 7-brane configuration (gauge group and matter spectrum). Hence the Sen limit should be implemented by scaling (some of) the $\mathsf{s}_\kappa$, i.e.\footnote{In some cases, a weak coupling limit is possible but it necessarily generates extra gauge groups (see~\cite{Esole:2012tf}).}
\begin{equation}\label{scaling}
\mathsf{s}_\kappa\rightarrow \epsilon^{n_\kappa} \mathsf{s}_\kappa \qquad\mbox{ with } \qquad n_\kappa=0,1,2 \:,
\end{equation}
 such that we realize the scaling \eqref{scaling_b} for the $b_i$'s.\footnote{In doing this, it can happen that $b_4$ has a leading term that scales with $\epsilon$ but has also a term that scales with $\epsilon^2$. The term that survives in the weak coupling limit is of course only the order $\epsilon$ term. This is what happens for example in the weak coupling limit of the Morrison-Park model (one massless \U1 with charge 1 and 2 states), as discussed in \cite{MayorgaPena:2017eda}.}
 Of course, the $\mathsf{s}_\kappa$ that are in $b_2$ should not scale.
If the F-theory model we started with has several gauge groups, at weak coupling $\Delta_E=b_4^2-b_2b_6$ will split into several components (when intersected with $\xi^2=b_2$).

\begin{figure}[t]
\begin{center}
  \includegraphics[scale=.4]{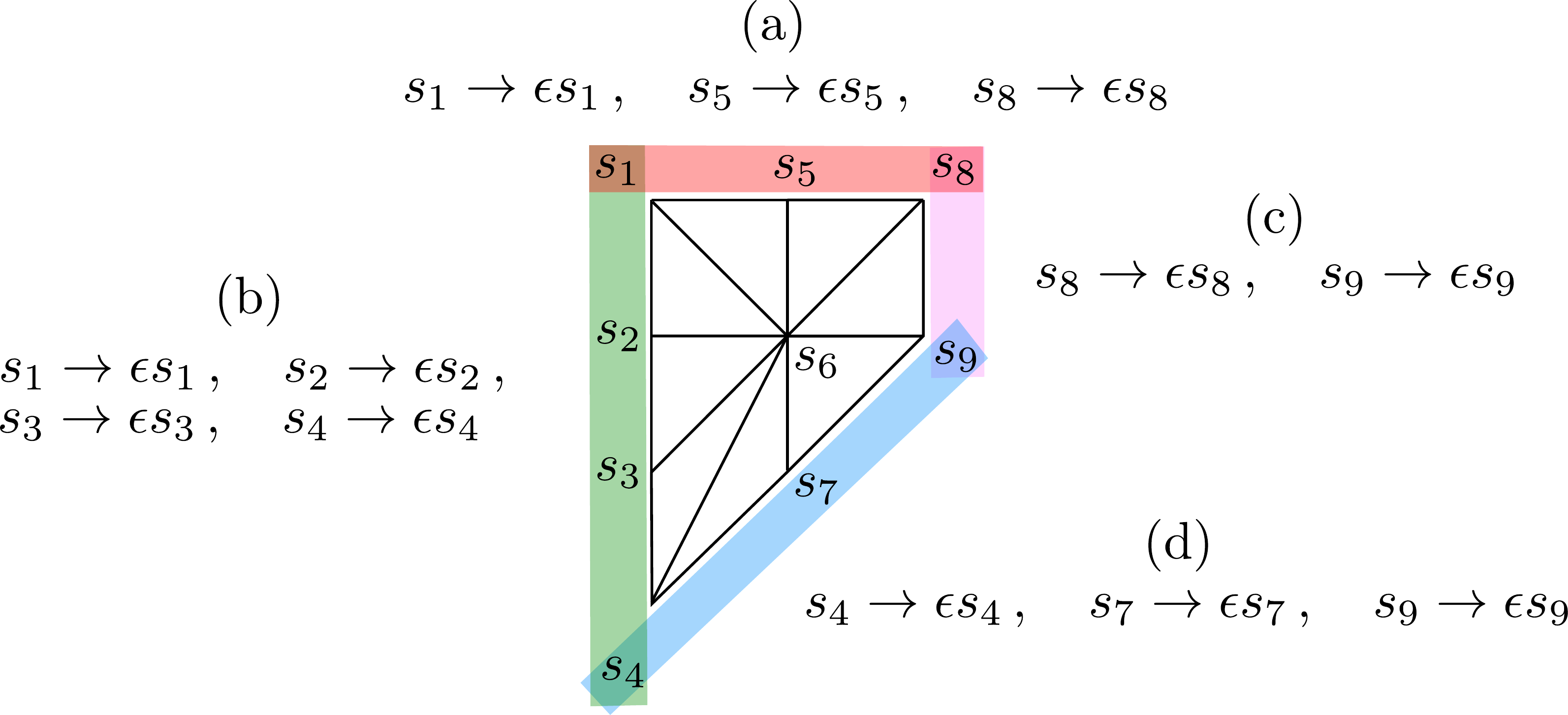}
\end{center}
\caption{\label{fig:wcls} The weak coupling limits for toric hypersurfaces can be constructed by scaling with $\epsilon$ all base sections along a line on the dual associated polytope while leaving the other base sections invariant. The polytope $F_{3}$ has four weak coupling limits associated to the facets of its dual polytope. There are two inequivalent weak couplings from the type IIB perspective as the limits (a) and (d), as well as (b) and (c) are equivalent because they produce the same D-brane configuration.
}
\end{figure}

There is a general observation regarding the weak coupling limits of toric hypersurface fibers. The consistent scalings leading to a perturbative type IIB model can be obatined with scalings of the form \eqref{scaling} with $n_\kappa=1$ for all the points $\kappa$ lying along a facet in the dual polytope, while leaving all other sections invariant under the scaling. This occurs for all of the 16 2D hypersurfaces considered in \cite{Klevers:2014bqa}. Take for example the polytope $F_3=dP_1$. One has then four weak coupling limits as indicated in Figure \ref{fig:wcls}. In reality, from the type IIB perspective such limits lead to only two inequivalent brane setups. 

\subsection{Massive and massless U(1)'s}\label{Sec:massiveVSmasslessU1}

In type IIB compactification, the \U1 symmetries live on the worldvolume of single D7-branes. If the D7-brane locus is invariant under the orientifold involution, the \U1 gauge boson is projected out; hence a \U1 symmetry is present if there is a pair of a D7-brane and its orientifold image. If the loci of these two branes are in the same homology class, then the \U1 gauge boson is massless. If the loci are in different homology classes, the gauge boson gets a mass through the ``geometric'' St\"uckelberg mechanism \cite{Ibanez:1998qp,Poppitz:1998dj,Aldazabal:2000dg}, by eating an axion. This axion comes from the reduction of the RR 2-form  $C_2$ along an odd two-form of the double cover CY.

If there are several massive \U1's in the compactification, some combinations of them may be massless. For example, if we have two massive \U1 gauge bosons and $h^{1,1}_-(X)=1$, then one combination of them will eat the only one axion, while the orthogonal combination will stay massless.

The fields living at the intersection of the D7-branes are charged under the \U1 symmetries. 
If we have two sets of brane/image-brane, $D7_1,D7_1'$ and $D7_2,D7_2'$, we have two sets of states: 1) at the intersection $D7_1 \cap D7_2$ with charges $(q_1,q_2)=(1,-1)$ and 2) at the intersection $D7_1 \cap D7_2'$ with charges $(Q_1,Q_2)=(1,1)$. If we have a set of brane/image-brane, $D7_1,D7_1'$ and an invariant D7-brane, we have states at the intersection $D7_1\cap D7_{\rm inv}$ with charge $Q_1=1$. If brane and image-brane intersect each other away from the orientifold plane, there are states at the intersection $D7_1\cap D7_1'$ with charge $Q_1=2$.

The charge corresponding to the massless gauge boson will be a linear combination of the charges of all the \U1's living on the D7-branes.

Let us see how one can find the massless $\U1$ generator in 6D compactifications (for the 4D case see \cite{Plauschinn:2008yd,Grimm:2011dj,Grimm:2011tb}). On the D7-brane worldvolume, the coupling that gives mass to the gauge bosons is given by:
\begin{equation}\label{CSgeomCoup}
 \int_{D7} C_6 \wedge \mathsf{F} \:,
\end{equation}
where $\mathsf{F} $ is the six-dimensional gauge boson field strength and $C_6$ is the dual of the RR two-form $C_2$.
The $C_6$ potential can be expanded as $C_6= \sum_\alpha c_4^\alpha \wedge \omega_\alpha^{(-)}$, with $\omega_\alpha^{(-)}\in H^{1,1}_-(X)$ and $c_4^\alpha$ six dimensional four-forms (dual in 6D to axionic scalar fields).
Plugging this expansion in \eqref{CSgeomCoup} we obtain 
\begin{equation}\label{CSgeomCoup2}
 \int_{D7} C_6 \wedge \mathsf{F} =\sum_\alpha \int_{\mathbb{R}^{1,5}} \mathsf{F}\wedge c_4^\alpha \int_{\mathcal{D}_{D7}} \omega_\alpha^{(-)}= 
 \sum_\alpha \int_{\mathbb{R}^{1,5}} \mathsf{F}\wedge c_4^\alpha \int_{X} \mathcal{D}_{D7}^{(-)}\wedge \omega_\alpha^{(-)}
\end{equation}
where $\mathcal{D}_{D7}= \mathcal{D}_{D7}^{(+)}+ \mathcal{D}_{D7}^{(-)}$ is the divisor wrapped by the D7-brane in the compact space and  $ \mathcal{D}_{D7}^{(+)}$ ($ \mathcal{D}_{D7}^{(-)}$) is the even (odd) component under the orientifold involution. The image-brane gives the same coupling term (it has opposite odd components and its field strength is $-\mathsf{F}$). An invariant brane does not have such a coupling, as its odd component is zero.

If we have $N$ massive $\U1$ branes, we will have the following term in the six dimensional effective action
\begin{equation}
\sum_{i=1}^N\sum_{\alpha=1}^{h^{1,1}_{-,{\rm eff}}} \int_{\mathbb{R}^{1,5}} n^i_\alpha\mathsf{F}_i\wedge c_4^\alpha  \qquad \mbox{with} \qquad n^i_\alpha = \int_{X} \mathcal{D}_{D7_i}^{(-)}\wedge \omega_\alpha^{(-)}
\end{equation}
where $h^{1,1}_{-,{\rm eff}}$ is the dimension of the subspace of $H^{1,1}_-(X)$ generated by $\mathcal{D}_{D7_i}^{(-)}$ ($i=1,...,N$)\footnote{The divisors $\mathcal{D}_{D7_i}^{(-)}$ are not independent of each other in general.} and $\omega_\alpha^{(-)}$ ($\alpha=1,...,h^{1,1}_{-,{\rm eff}}$) are elements of a basis in $H^{1,1}_-(X)$.

If $N=h^{1,1}_{-,{\rm eff}}$, all the \U1 gauge bosons get a mass by St\"uckelberg mechanism \cite{Ibanez:1998qp,Poppitz:1998dj,Aldazabal:2000dg}. On the other hand, if $N>h^{1,1}_{-,{\rm eff}}$ there are $N-h^{1,1}_{-,{\rm eff}}$ massless combinations.


\subsection{Minimal models with high charges}
\label{sec:minimal}

We want to construct a model in type IIB with one massless $\U1$ and with states that have high charge under this \U1.

The easiest way to realize a massless $\U1$ is to take a pair of a D7-brane and its image in the same homology class (see \cite{Grimm:2010ez} for an F-theory realization and \cite{Braun:2011zm,Braun:2014pva} for the weak coupling limit). If there are no other branes, there will be only a state at the intersection $D7\cap D7'$ that will have unit charge. On the other hand, if there is another (invariant)\footnote{If there is another pair, it must be massive, otherwise we would have two massless \U1's.} 
brane, there  will be a state with charge $1$ at $D7\cap D7_{\rm inv}$ and a state with charge $2$ at the intersection of the D7-brane with its image, away from the orientifold locus, i.e. at  $(D7\cap D7')\setminus(D7\cap O7)$ (see \cite{Morrison:2012ei} for an F-theory realization and \cite{MayorgaPena:2017eda} for its weak coupling limit). 

To obtain \U1 models with charges higher than $2$, one needs to introduce massive \U1 D7-branes in a CY double cover with $h^{1,1}\neq 0$. Since we want to end up with only one massless gauge boson, we need that the number of massive \U1 D7-branes is one unit bigger than the number of axions to be eaten, that is equal to $h^{1,1}_{-,{\rm eff}}$.
The minimal choice is $h^{1,1}_{-,\rm eff}=1$. We will see that under this assumption we will construct the weak coupling limit of all the high charge F-theory models known so far. We will then consider the case when the double cover CY is given by the equation 
$$
\xi^2=\frac{s_6^2}{4}-s_Ls_R \:.
$$
As explained in Section~\ref{Sec:DoubCov}, for generic sections $s_6,s_L,s_R$ of the corresponding line bundles on $B$, this space has an odd 2-form dual to the divisor $D_-$, whose class can be represented by the difference $D_p-D_p'$, where $D_p,D_p'$ are given in \eqref{DivInDC}.

\begin{figure}[h]
\centering
\begin{minipage}{.33\textwidth}
 \centering
  \setlength{\unitlength}{0.2\textwidth}
  \begin{picture}(4,5)
    \put(0.3,0){\includegraphics[scale=.33]{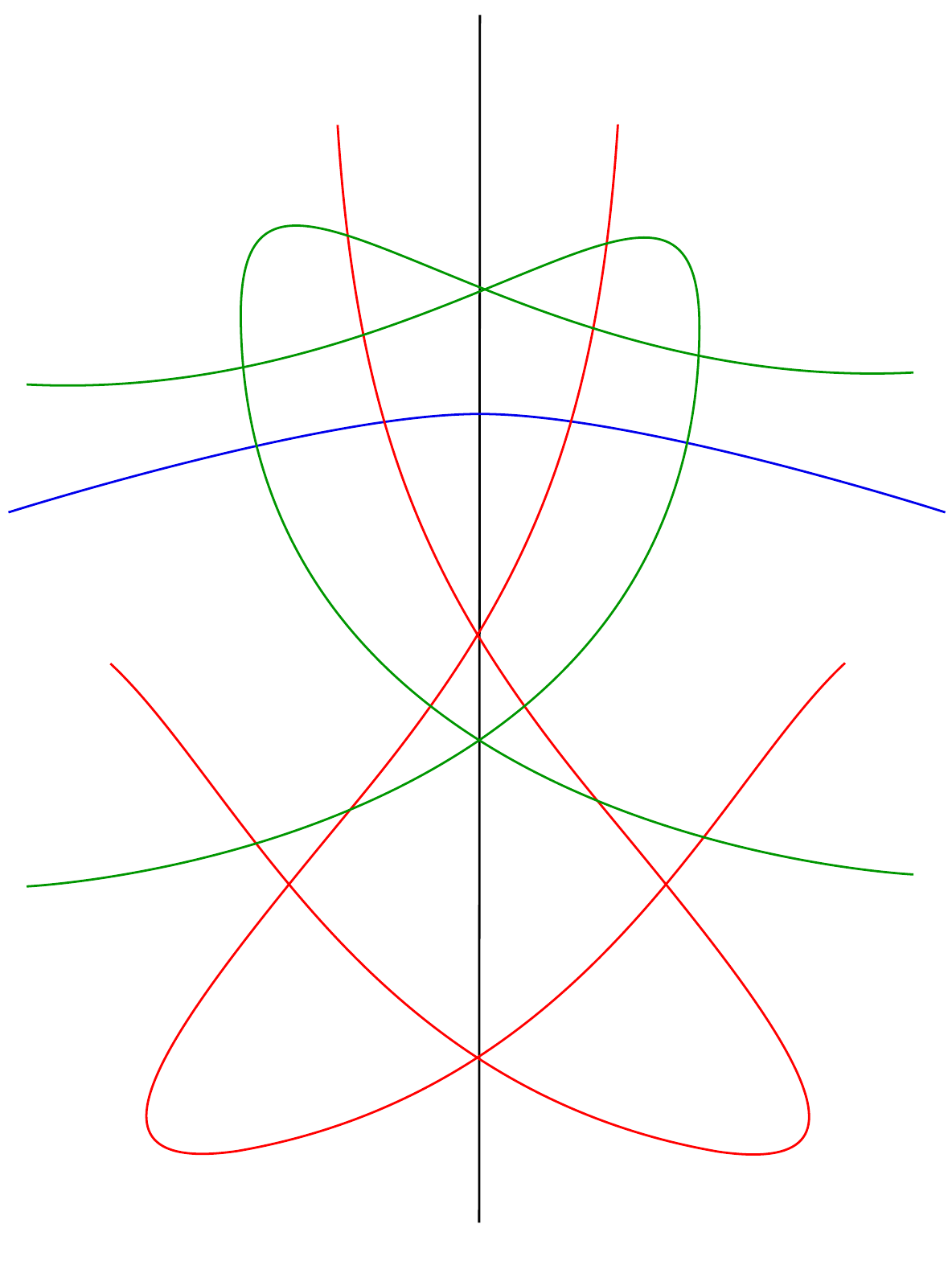}}
        \put(0.3,2.55){\small{\color{blue} $D7_{\rm inv}$}}
       \put(0.45,3.7){\small{\color{verde} $D7_{1}^\prime$}}
    \put(3.4,3.7){\small{\color{verde} $D7_1$}}
     \put(0.18,0.8){\small{\color{red} $D7_2^\prime$}}
   \put(3.5,0.8){\small{\color{red} $D7_2$}}
   \put(2.25,4.6){\small $O7$}
  \end{picture}\\
  (i)
\end{minipage}%
\begin{minipage}{.33\textwidth}
 \centering
  \setlength{\unitlength}{0.2\textwidth}
  \begin{picture}(4,5)
    \put(0.3,0){\includegraphics[scale=.33]{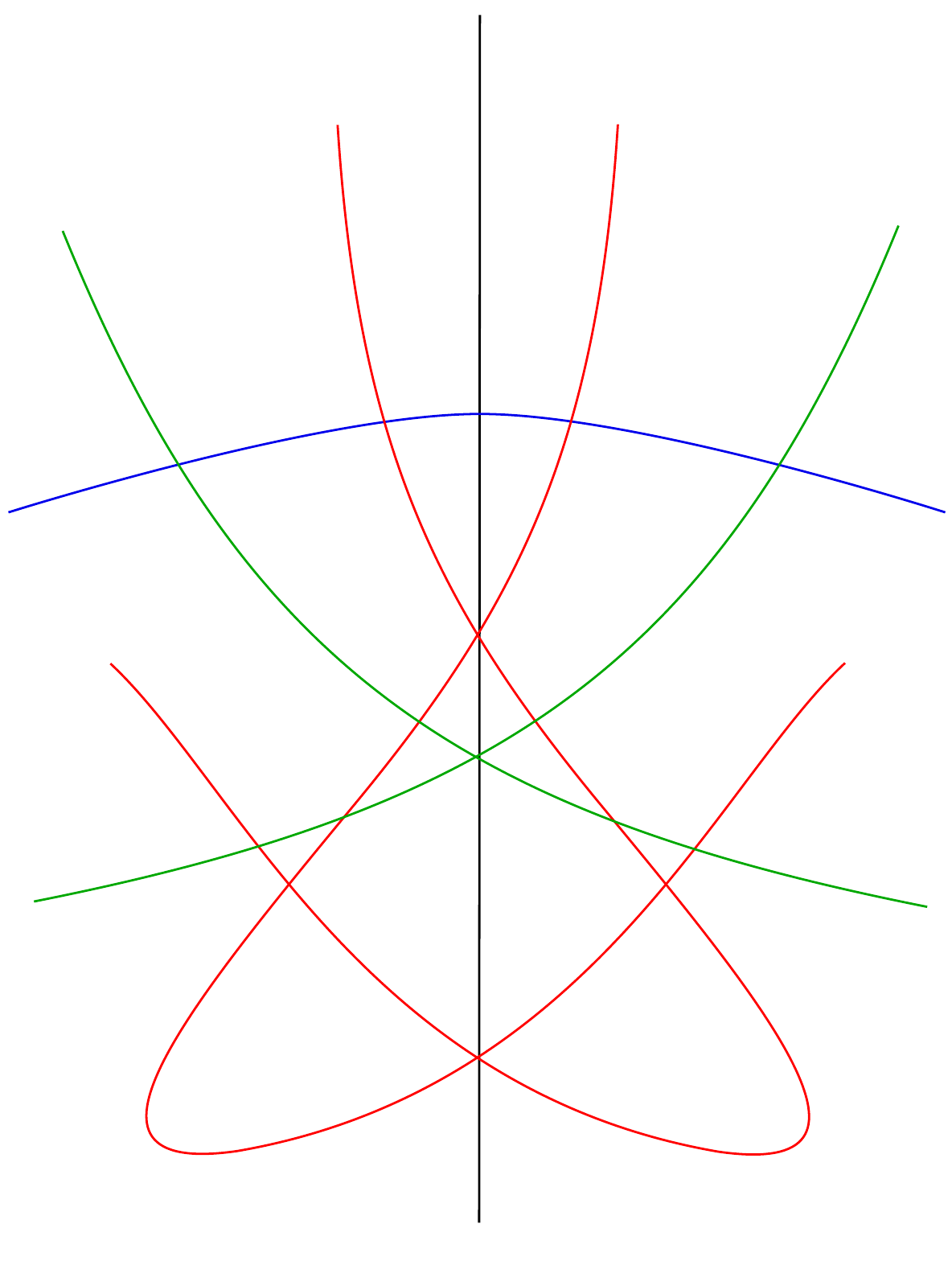}}
  \put(0.3,2.55){\small{\color{blue} $D7_{\rm inv}$}}
        \put(0.8,3.7){\small{\color{verde} $D7_{1}^\prime$}}
    \put(3.0,3.7){\small{\color{verde} $D7_1$}}
   \put(0.18,0.8){\small{\color{red} $D7_2^\prime$}}
   \put(3.5,0.8){\small{\color{red} $D7_2$}}
   \put(2.25,4.6){\small $O7$}
  \end{picture}\\
  (ii)
\end{minipage}
\caption{\label{fig:Charge3456} (i) When branes and image branes intersect away from the orientifold it is possible to get up to charge 4 (with $a=2$, $b=1$) or up to charge 6 (with $a=3$, $b=2$). (ii) If $D7_1$ and $D7_1^\prime$ intersect only on top of the $O7$, one can only get up to charge 3 (with $a=2$, $b=1$) or up to charge 5 (with $a=3$, $b=2$).}
\end{figure}
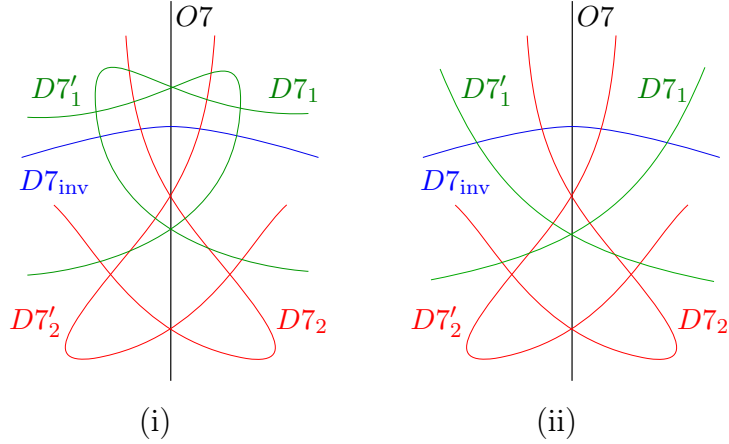

In this situation, we obtain one massless \U1 from two would-be geometrically massive  \U1's: one combination of them will eat the axion associated to $D_-$ whereas the orthogonal combination remains a massless~\U1. In type IIB we then have two pairs of massive D7 brane/image-brane, plus a possible invariant D7-brane (see Figure~\ref{fig:Charge3456}). Let's assume that the odd brane divisor classes satisfy 
\begin{equation}
 \mathcal{D}_{D7_1}^{(-)}=b D_- \qquad\mbox{and}\qquad \mathcal{D}_{D7_2}^{(-)}=-aD_-\:;
 \end{equation} 
then the six-dimensional coupling will be (with the normalization $\int_X\omega^{(-)}\wedge D_- =1$)
\begin{equation}
 \int_{\mathbb{R}^{1,5}} b \,\mathsf{F}_1 \wedge c_4 -   \int_{\mathbb{R}^{1,5}} a\, \mathsf{F}_2 \wedge c_4 =  \int_{\mathbb{R}^{1,5}} (b\, \mathsf{F}_1-a \,\mathsf{F}_2) \wedge c_4  \:.
\end{equation}
Hence the combination $b\,\mathsf{F}_1-a\, \mathsf{F}_2$ is massive and the orthogonal combination $a\,\mathsf{F}_1+b \,\mathsf{F}_2$ is massless. In particular, the charge associated with the massless \U1 will be a combination of the two massive \U1 charges:
\begin{equation}\label{masslessU1chargeGEN}
Q = \frac{1}{\kappa} \left( a \,Q_1 + b\, Q_2 \right) \:,
\end{equation}
where $a,b$ are integers and $\kappa$ is the greatest common divisor of the \U1 charges $a Q_1+b Q_2$ of the states in the configuration.
In Table~\ref{tab:2branes1inv} we report the massless \U1 charge for each state in the configuration we have chosen.
\begin{table}[h]
\begin{center}
\renewcommand{\arraystretch}{1.4}
{
\begin{tabular}{|c|c|c|c|c|c|c|}\cline{2-7}
\multicolumn{1}{c|}{} & $D7_1\cap D7_2$ &  $D7_1\cap D7_2'$ & $D7_1\cap D7_1'$ & $D7_2\cap D7_2'$  & $D7_1\cap D7_{\rm inv}$  & $D7_2\cap D7_{\rm inv}$  \\ \hline
$(Q_1,Q_2)$ & $(1,-1)$ & $(1,1)$ & $(2,0)$ & $(0,2)$& $(1,0)$  & $(0,1)$   \\ \hline\hline
$Q$ & $(a-b)/\kappa$ & $(a+b)/\kappa$ & $2a/\kappa$ & $2b/\kappa$ & $a/\kappa$ & $b/\kappa$ \\ \hline
\end{tabular}
}
\caption{\label{tab:2branes1inv} \U1 charges for generic $a,b$ (see \eqref{masslessU1chargeGEN}).} 
\end{center}
\end{table}

We will consider only configurations in which
the different intersections  provide matter with charges taking all the integral values between $1$ and  $Q_{\rm max}$, where $Q_{\rm max}$ is the maximal charge realized in the model.
In this way we will automatically satisfy the completeness conjecture. It in fact states that an effective field theory with a \U1 gauge symmetry is only consistent with quantum gravity provided that all charges $Q\in \mathbb{Z}$ appear at some level in the mass spectrum \cite{Polchinski:2003bq,Banks:2010zn}.%
\footnote{While this observation is not entirely restricitive for the massless spectrum of our interest, we rely on the generic observation that for all F-theory models with \U1 symmetries constructed so far, the structure of the elliptic fiber allows for singularities at base codimension two, such that all charges between 1 and a certain maximum charge $Q_{\rm max}$ are possible. However, it could happen that over a specific base some of the matter loci are empty and then there are no fields with a given charge $Q_i$ ($1 \leq Q_i\leq Q_{\rm max}$) in the massless spectum (see \cite{Taylor:2018khc} for further details and examples of this phenomenon). This can also happen for some realization of the models we  construct in this paper.}

Notice that in Table~\ref{tab:2branes1inv}  there are six different states. Hence if we want that all the values of $Q$ are filled up to $Q_{\rm max}$, then we can only construct models {\bf with the highest charge up to 6}. Hence with the chosen configuration we have models with $Q_{\rm max}=3,4,5,6$:
\begin{description}
\item {\bf Charge 3 (I)}: $a=2$, $b=1$, $\kappa=1$ and $D7_1\cap D7_1' \setminus D7_1\cap O7=\emptyset$ with the presence of an invariant brane~$D7_{\rm inv}$.
\item {\bf Charge 3 (II)}: $a=3$, $b=1$, $\kappa=2$, non empty $D7_1\cap D7_1' \setminus D7_1\cap O7$ and no invariant brane. 
\item {\bf Charge 3 (III)}: $a=5$, $b=1$, $\kappa=2$, $D7_1\cap D7_1' \setminus D7_1\cap O7=\emptyset$ and no invariant brane. 
\item {\bf Charge 3 (IV)}: $a=6$, $b=2$, $\kappa=4$, non empty $D7_1\cap D7_1' \setminus D7_1\cap O7$ and no invariant brane. 
\item {\bf Charge 4 (I)}: $a=2$, $b=1$, $\kappa=1$ and non-empty $D7_1\cap D7_1'\setminus D7_1\cap O7$ with the presence of an invariant brane~$D7_{\rm inv}$.
\item {\bf Charge 4 (II)}: $a=4$, $b=2$, $\kappa=2$, non-empty $D7_1\cap D7_1'\setminus D7_1\cap O7$ with or without an invariant brane. 
\item {\bf Charge 5 (I)}: $a=3$, $b=2$, $\kappa=1$  and $D7_1\cap D7_1'\setminus D7_1\cap O7=\emptyset$ with the presence of an invariant brane~$D7_{\rm inv}$.
\item {\bf Charge 5 (II)}: $a=4$, $b=1$, $\kappa=1$  and $D7_1\cap D7_1'\setminus D7_1\cap O7=\emptyset$ with the presence of an invariant brane~$D7_{\rm inv}$.
\item {\bf Charge 6}: $a=3$, $b=2$ and non-empty $D7_1\cap D7_1'\setminus D7_1\cap O7$.
\end{description}
In principle,  in the cases with $a=2$ and $b=1$ the invariant brane is not necessary to realize all charges up to $Q_{\rm max}$, as the charges of $D7_i\cap D7_i'$ are already realized in other intersections. We will see in the following that the invariant brane is necessary when $a+b$ is odd.  We will also see at the end of Section~\ref{Sec:6DIIBCompactif} why we have secretly considered the bound $a+b\leq 8$.
All the values of $a,b$ satisfying this bound and that we did not consider correspond to models where some of the charge values is not populated by actual massless states.
We do not consider models with $a=b$, that always give states with zero charge. 


\subsection{6D compactification and D7-brane setup}\label{Sec:6DIIBCompactif}

We consider 6D type IIB compactifications  on a $K3$ surface $X$ with an orientifold projection.\footnote{For some aspects of type IIB 6D models related to F-theory constructions see \cite{Braun:2009bh}.} Holomorphic involutions on $K3$ were classified by Nikulin \cite{Nikulin} in terms of three integer parameters 
$(r,\mathrm{a},\delta)$. In particular $r$ is the number of $K3$ two-cycles that are even under the involution, i.e. $b_2^+(X)=r$ and $b_2^-(X)=22-r$ (remember that $b_2(K3)=22$).
The fixed point locus is always given (except for two special cases) by the disjoint union of a genus $g$ curve and $k$ spheres, that have the following expressions in terms of $r$ and $\mathrm{a}$:
\begin{equation}
g = \frac12(22-r-\mathrm{a}), \qquad k=\frac12 (r-\mathrm{a}) \:.
\end{equation}
In our case, the fixed point locus $\xi=0$ that lives in the homology class $c_1(\L)=\bar{K}_B$ (where $\bar{K}_B$ is the anticanonical class of the quotient $B$, pulled back to $X$). We consider cases where we have only one connected O7-plane. This means that $k=0$, i.e. $r=\mathrm{a}$.\footnote{When $k\neq 0$, the O7-planes wrap rigid two-sphere in $K3$ with gauge group $SO(8)$. The corresponding F-theory lift will have 
Non-Higgsable Clusters with non-abelian $D_4$ singularities. Since we are interested in abelian gauge groups, we consider only involutions with $k=0$.} The genus of the O7-locus is
\begin{equation}
 g = \frac12(2-\chi_{O7})=\frac12\left(2 - \int_{O7} c_1(O7) \right)= 1+ \frac12 \bar{K}_B^2\:.
\end{equation}
We can then derive the relations
\begin{equation}
b_2^+(X) \,=\,r \,=\, 10 - \frac12 \bar{K}_B^2 \qquad\mbox{ and }\qquad b_2^-(X)\,=\, 22-r \,=\, 12 + \frac12 \bar{K}_B^2  \:.
\end{equation}

The 6D effective theory has $\mathcal{N}=1$ supersymmetry. The low energy spectrum will be made up by the gravity multiplet, $V$ vector multiplets, $T$ tensor multiplets and $H$ hypermultiplets. The number of vectors is given by the number of non-invariant D7-branes (if the D7-brane worldvolume is invariant, the corresponding gauge field is projected out by the orientifold). The number of tensors is given by $b_2^+(X)$; considering that one tensor sits in the gravity multiplet, we have 
\begin{equation}
T \,=\, b_2^+(X) - 1 \, = \, 9 - \frac12 \bar{K}_B^2\:
\end{equation} 
tensor multiplets, that includes also $b_2^+(X) - 1$ K\"ahler moduli. There are then $2b_2^-(X)$ further complex scalars that organize in $H_{\rm bulk}=b_2^-(X)= 12 + \frac12 \bar{K}_B^2 $ hypermultiplets:
\begin{itemize}
\item $1$ axio-dilaton $C_0+i\,e^{-\phi}$;
\item $1$ volume modulus complexified by $C_4$ along the volume form of $X$;
\item $b_2^-(X)-2$ complex structure moduli;
\item $b_2^-(X)$ complex scalars coming from reducing $B_2+iC_2$ along the odd two-forms.
\end{itemize}

The open string sector introduces further hypermultiplets: there are neutral hypermultiplet that include the open string moduli (deformations of the D7-branes and the Wilson lines) and the charged hypermultiplets living at the intersection of the D7-branes.

In all the models we consider in this paper, 
we will consider configurations with two pairs of (massive) brane/image-brane, say $D7_1/D7_1'$ and $D7_2/D7_2'$, and (in most cases) an invariant brane $D7_{\rm inv}$. 

Let us begin considering the pairs of brane/image-brane.
Given the consideration in Section~\ref{sec:minimal}, the divisors wrapped by the branes are in the classes
\begin{eqnarray}
\label{D71classes} \left[\mathcal{D}_{D7_1}\right] = b [D_{L,+}] + [x_1] &\qquad &  [\mathcal{D}_{D7_1'}] = b [D_{L,-}] + [x_1]\\
\label{D72classes}\left[\mathcal{D}_{D7_2}\right] = a [D_{L,-}] + [y_1] &\qquad &  [\mathcal{D}_{D7_2'}] = a [D_{L,+}] + [y_1]
\end{eqnarray}
where the corresponding loci are given by (see Section~\ref{Sec:DoubCov})
\begin{eqnarray}
\label{locusD71D71p} \mathcal{D}_{D7_1}: \,\,\, \tilde{\mathrm{M}}_{\rm red}^{\otimes b} \cdot \vec{x}=0  &\qquad&  \mathcal{D}_{D7_1'}: \,\,\, \mathrm{M}_{\rm red}^{\otimes b} \cdot \vec{x}=0 \\
\label{locusD72D72p} \mathcal{D}_{D7_2}: \,\,\, \mathrm{M}_{\rm red}^{\otimes a} \cdot \vec{y}=0   &\qquad&  \mathcal{D}_{D7_2'}: \,\,\, \tilde{\mathrm{M}}_{\rm red}^{\otimes a} \cdot \vec{y}=0 
\end{eqnarray}
with $\vec{x}=(x_1,...,x_{a+1})$ and $\vec{y}=(y_1,...,y_{b+1})$, and ${\rm M}_{\rm red}^{\otimes k}$ the $(k+1)\times (k+1)$ non trivial block of $\mathrm{M}^{\otimes k}$.

The number of open string moduli for each pair is given by the genus of the surface $\mathcal{D}_{D7_i}$,~i.e.
\begin{equation}\label{genusD7i}
g_{D7_i} = \frac12 \left( 2 + \int_X \mathcal{D}_{D7_i}^2 \right) \:.
\end{equation}
Plugging \eqref{D71classes} and \eqref{D72classes} into \eqref{genusD7i} and using the relations \eqref{PDLp}, \eqref{DLpDLm} and \eqref{DLpSquare} one obtains
\begin{eqnarray}
g_{D7_1} &=& 1 + \tfrac{b^2}{4} [s_L]\cdot ([s_L] - \bar{K}_B) + \tfrac{b}{2} [s_L]\cdot [x_1] + \tfrac12 [x_1]^2\:,  \\
g_{D7_2} &=& 1 + \tfrac{a^2}{4} [s_L]\cdot ([s_L] - \bar{K}_B) + \tfrac{a}{2} [s_L]\cdot [y_1] + \tfrac12 [y_1]^2\:.
\end{eqnarray}
The number of charged hypermultiplets at the inersection of branes coming from different pairs is given by the intersection numbers $[\mathcal{D}_{D7_1}]\cdot [\mathcal{D}_{D7_2}]$ and $[\mathcal{D}_{D7_1}]\cdot [\mathcal{D}_{D7_2'}]$:
\begin{eqnarray}
N_{D7_1\cap D7_2} &=& \frac{ab}{2}[s_L]\cdot \bar{K}_B + \frac12 [s_L]\cdot (a[x_1]+b[y_1])+[x_1]\cdot [y_1] \:, \\
N_{D7_1\cap D7_2'} &=& \frac{ab}{2}[s_L]\cdot ([s_L]- \bar{K}_B) + \frac12 [s_L]\cdot (a[x_1]+b[y_1])+[x_1]\cdot [y_1] \:. 
\end{eqnarray}
The number of states at the intersection of a brane with its image is instead given by~$\frac12 ( [\mathcal{D}_{D7_i}]\cdot [\mathcal{D}_{D7_i'}] - [O7]\cdot [\mathcal{D}_{D7_i}])$:
\begin{eqnarray}
N_{D7_1\cap D7_1'} &=& \frac{b(b-1)}{4}[s_L]\cdot \bar{K}_B + \frac{b}{2} [s_L]\cdot [x_1] -\frac12 \bar{K}\cdot[x_1]+ \frac12 [x_1]^2\:, \\
N_{D7_2\cap D7_2'} &=& \frac{a(a-1)}{4}[s_L]\cdot \bar{K}_B + \frac{a}{2} [s_L]\cdot [y_1] -\frac12 \bar{K}\cdot[y_1]+ \frac12 [y_1]^2 \:.
\end{eqnarray}

The invariant brane can be obtained by recombining a pair of brane/image-brane. The recombination can be described by a Higgs mechanism: a field living at the intersection of the brane with its image gets a non-zero vev, the vector multiplet living on the D7-brane gets a non-zero mass by eating one of the charged hypermultiplets. The number of open string moduli hypermultiplets of the invariant brane is then given by
\begin{equation}
 n_{\rm inv} = g_{D7} + N_{D7\cap D7'} - 1 \:.
\end{equation}
In what follows we will always consider invariant branes obtained by recombining a pair of brane/image-brane wrapping a divisors in the classes $[D_{L,\pm}]+[w_2]$, i.e. it will wrap the locus 
\begin{equation}\label{invD7ord1}
P_{D7_{\rm inv}} = s_Rw_1^2 -s_6w_1w_2+s_Lw_2^2 + 4\left( \frac{s_6^2}{4}-s_Ls_R \right) w_3  \:
\end{equation}
in the class $[\mathcal{D}_{D7_{\rm inv}}] = [s_L] + 2 [w_2]=2\bar{K}_B+[w_3]$.
We then have 
\begin{equation}
 n_{\rm inv} = \tfrac14 [s_L]\cdot ([s_L] - \bar{K}_B)  +[s_L]\cdot [w_2] -\tfrac12 \bar{K}\cdot[w_2]+ [w_2]^2 \:.
\end{equation}
The invariant brane intersects the branes $D7_1$ and $D7_2$, giving respectively $[\mathcal{D}_{D7_{\rm inv}}]\cdot [\mathcal{D}_{D7_{1}}]$ and $[\mathcal{D}_{D7_{\rm inv}}]\cdot [\mathcal{D}_{D7_{2}}]$ hypermultiplets:
\begin{eqnarray}
N_{D7_{\rm inv}\cap D7_1} &=& \frac{b}{2}[s_L]^2 +b [s_L]\cdot [w_2] + [s_L]\cdot [x_1] + 2 [x_1]\cdot [w_2] \:, \\
N_{D7_{\rm inv}\cap D7_2} &=& \frac{a}{2}[s_L]^2 +a [s_L]\cdot [w_2] + [s_L]\cdot [y_1] + 2 [y_1]\cdot [w_2] \:.
\end{eqnarray}

If we sum the $H_{\rm bulk}$ hypermultiplets and the open string hypermultiplets, we obtain the total number of hypermultiplets. In both cases, with and without invariant brane we obtain
\begin{equation}
   H = 14+\frac{29}{2}\bar{K}^2 = V + 273 - 29T
\end{equation}
i.e. this configuration satisfies the gravitational anomaly cancellation condition, as it should be for a D-brane configuration with all D-brane charges canceled. The same occurs for the mixed $\U1$-gravitational as well as the pure $\U1$ anomaly. 
The details of the anomaly computation can be found in Appendix~\ref{App:AnomalyCancel}.

We finish this section with the D7-brane tadpole cancellation condition. In the case when $D7_{\rm}$ is present, we have 
\begin{equation}
[\mathcal{D}_{D7_1}]+[\mathcal{D}_{D7_1'}]+[\mathcal{D}_{D7_2}]+[\mathcal{D}_{D7_2'}]+  [\mathcal{D}_{D7_{\rm inv}}]  = 8[O7] \:.
\end{equation}
Plugging \eqref{D71classes}, \eqref{D72classes}, $[\mathcal{D}_{D7_{\rm inv}}] = [s_L] + 2 [w_2]$ and $[O7]=\bar{K}_B$ into this expression we obtain
\begin{equation}\label{xywD7tadpt}
 (a+b)[s_L] + 2[x_1] + 2[y_1] + 2\bar{K}_B+[w_3] = 8\bar{K}_B \:.
\end{equation}
We now use the condition that $a[s_L]+[x_1]=a[s_6]+[x_{a+1}]$, as it can be seen from \eqref{locusD71D71p}. Analogously one has $b[s_L]+[y_1]=b[s_6]+[y_{b+1}]$. 
Moreover, remember that $[s_6]=\bar{K}_B$. We can then write the equation \eqref{xywD7tadpt} in the form
\begin{equation}\label{TadpConstrab}
 (6-a-b)\bar{K}_B =  [x_1]+[x_{a+1}]+[y_1]+[y_{b+1}]+[w_3]   \:.
\end{equation}
For connected loci of D7-branes (except the case in which $[D7]=[D_{L,\pm}]$ or $[D7]=[D_{R,\pm}]$), the classes $[x_i]$, $[y_j]$ and $[w_\ell]$ are effective. $\bar{K}_B$ is effective as well (it is the class of the orientifold plane).
The equation \eqref{TadpConstrab} then constrains $a$ and $b$ to be bounded as $a+b\leq 6$ when there is an invariant brane like the one described in Eq. \eqref{invD7ord1}. If there is no invariant brane, then (by analogous computations) the bound becomes $a+b\leq 8$, as anticipated in the previous section.

\section{Charge 3 and 4 models and their F-theory lift}\label{Sec:charge34}

\subsection{Type IIB models with maximal charge 3 or 4}

In this section we will consider  models with charge 3 (I) and (II) and model with charge~4~(I). As we will see the first two come from the Sen limit of the same F-theory model. 

\subsubsection*{Charge 3 (I) and charge 4 (I) models}
We start with
$a=2$ and $b=1$ and an invariant brane, that realizes the charge 3 (I) and charge 4 (I) models. The values of $a$ and $b$ imply that
\begin{equation}
  \mathcal{D}_{D7_2} -  \mathcal{D}_{D7_2'} = -2( \mathcal{D}_{D7_1} -  \mathcal{D}_{D7_1'}) \:,
\end{equation}
where $\mathcal{D}_{D7}$ is the divisor class wrapped by the $D7$ brane. 
The $D7_1$ locus in the quotient is $P_{D7_1}=0$ where $P_{D7_1}$ is given by \eqref{DBp1p2}, i.e.
 \begin{equation}\label{EqD7p1p2}
 P_{D7_1} = s_Rp_1^2 -s_6p_1p_2+s_Lp_2^2 \:,
 \end{equation}
 while the $D7_2$ locus in the quotient is $P_{D7_2}=0$ where $P_{D7_2}$ is given by \eqref{DB2inquot}, i.e. 
\begin{equation}\label{EqD7q1q2}
 P_{D7_2} = \left( s_Lq_1-s_Rq_3 \right)^2  - \left( s_Lq_2-s_6q_3 \right)\left( s_6q_1-s_Rq_2 \right) \:.
 \end{equation}
The only difference between $Q_{\rm max}=3$ and $Q_{\rm max}=4$ models is that in the first  case the intersection $D7_1\cap D7_1'$ is empty outside the O7-plane locus. This means that, while for the $Q_{\rm max}=4$ model the vector $(p_1,p_2)$ is generic, for the $Q_{\rm max}=3$ model it is given by either $(1,0)$ or $(0,1)$.

The invariant brane must satisfy the constraint that it has double intersection with the orientifold plane $\xi=0$, i.e. on top of the O7-plane it must split as a brane/image-brane pair \cite{Collinucci:2008pf,Braun:2008ua}.  In particular, we require that the polynomial $\Delta_E$ (where $\Delta_E=0$ is the full D7-brane locus in \eqref{DB2inquot}) reduces to a square on top of the O7-plane $b_2=0$ \cite{Collinucci:2008pf,Braun:2008ua}. Since $s_{L},s_R,s_6$ appear linearly in $P_{D7_1}$ (for generic $p_1,p_2$) and quadratically in $P_{D7_2}$ (for generic $q_1,q_2,q_3$), then they should appear to an odd power in $P_{D7_{\rm inv}}$, up to a polynomial that vanishes if $b_2=0$. 
If we make the minimal choice (in which $s_6$, $s_L$ and $s_R$ appear linearly), we have an invariant brane wrapping the locus \eqref{invD7ord1}.\footnote{With a different choice we would have added more coefficients $r_i$, with a higher restriction on their degrees and the degrees of the coefficients $p_j,q_k$. This choice will give the most generic such situation; specializing these coefficients, one can realize the configuration with higher powers of $s_{L,6,R}$.}

The full D7-brane locus is then $\Delta_E=0$, where
\begin{equation}
\Delta_E = P_{D7_1}  P_{D7_2}  P_{D7_{\rm inv}} \:.
\end{equation}

The charges of the states and their locations are reported in Table~\ref{tab:Ch34IIB}.
We notice that in type IIB there are two loci corresponding to charge $2$ and two corresponding to charge $1$. As we shortly see, loci with the same charge recombine away from weak coupling, giving a unique locus with charge $1$ and a unique locus with charge $2$ in F-theory.
\begin{table}[h!]
\begin{center}
\renewcommand{\arraystretch}{1.4}
{\footnotesize
\begin{tabular}{|c|c|c|c|c|c|c|}\cline{2-7}
\multicolumn{1}{c|}{} & $D7_1\cap D7_2$ &  $D7_1\cap D7_2'$ & $D7_1\cap D7_1'$ & $D7_2\cap D7_2'$  & $D7_1\cap D7_{\rm inv}$  & $D7_2\cap D7_{\rm inv}$  \\ \hline
$(Q_1,Q_2)$ & $(1,-1)$ & $(1,1)$ & $(2,0)$ & $(0,2)$& $(1,0)$  & $(0,1)$   \\ \hline\hline
$Q$ & $1$ & $3$ & $ 4 $ & $2$ & $2$ & $1$ \\ \hline
\end{tabular}
}
\caption{\label{tab:Ch34IIB} \U1 charges for charge 3 (I) and charge 4 (I) models in type IIB. The state at $D7_1\cap D7_1'$ is present only in charge 4 models.} 
\end{center}
\end{table}

\subsubsection*{Charge 3 (II) model}

We now consider the charge 3 (II) model, i.e. $a=3$, $b=1$. In this case there is no invariant brane.  We then have
\begin{equation}
  \mathcal{D}_{D7_2} -  \mathcal{D}_{D7_2'} = -3( \mathcal{D}_{D7_1} -  \mathcal{D}_{D7_1'}) \:.
\end{equation}
The $D7_1$ locus in the quotient $B$ is $P_{D7_1}=0$ where $P_{D7_1}$ is again given by \eqref{DBp1p2}, i.e.
 \begin{equation}\label{EqD7q1q2II}
 P_{D7_1} = s_Rp_1^2 -s_6p_1p_2+s_Lp_2^2\:.
 \end{equation}
The $D7_2$ locus in the quotient is $P_{D7_2}=0$ where now $P_{D7_2}$ is given by \eqref{DB2inquot3} , i.e.
\begin{align}
\begin{split}\label{EqD7r1r2II}
 P_{D7_2} =&\, r_1^2 s_L^3 - r_1 s_6 (r_4 s_6^2 + s_L (-r_3 s_6 + r_2 s_L)) + 
 r_1 s_L (3 r_4 s_6 - 2 r_3 s_L) s_R \\
 &+ 
 s_R (r_2^2 s_L^2 + s_R (-r_3 r_4 s_6 + r_3^2 s_L + r_4^2 s_R) + 
    r_2 (r_4 s_6^2 - r_3 s_6 s_L - 2 r_4 s_L s_R))\,
    \end{split}\:.
 \end{align}
The full D7-brane locus is then $\Delta_E=0$, where
\begin{equation}
\Delta_E = P_{D7_1}  P_{D7_2} \:.
\end{equation}
The charges of the states and their locations are reported in Table~\ref{tab:Ch3II_IIB}.We notice that in type IIB there are two loci corresponding to charge $1$. 
These will recombine away from weak coupling.
\begin{table}[h!]
\begin{center}
\renewcommand{\arraystretch}{1.4}
{\footnotesize
\begin{tabular}{|c|c|c|c|c|}\cline{2-5}
\multicolumn{1}{c|}{} & $D7_1\cap D7_2$ &  $D7_1\cap D7_2'$ & $D7_1\cap D7_1'$ & $D7_2\cap D7_2'$    \\ \hline
$(Q_1,Q_2)$ & $(1,-1)$ & $(1,1)$ & $(2,0)$ & $(0,2)$  \\ \hline\hline
$Q$ & $1$ & $2$ & $ 3 $ & $1$  \\ \hline
\end{tabular}
}
\caption{\label{tab:Ch3II_IIB} \U1 charges for charge 3 (II) models in type IIB.} 
\end{center}
\end{table}


\subsection{Charge 3 type IIB model from F-theory}

In F-theory the charge three model was described in \cite{Klevers:2014bqa}. The weak coupling limit was performed in \cite{MayorgaPena:2017eda}.
We summarize the result here.

\begin{figure}[H]
\centering
\begin{minipage}{.56\textwidth}
  \centering
  \includegraphics[scale=.4]{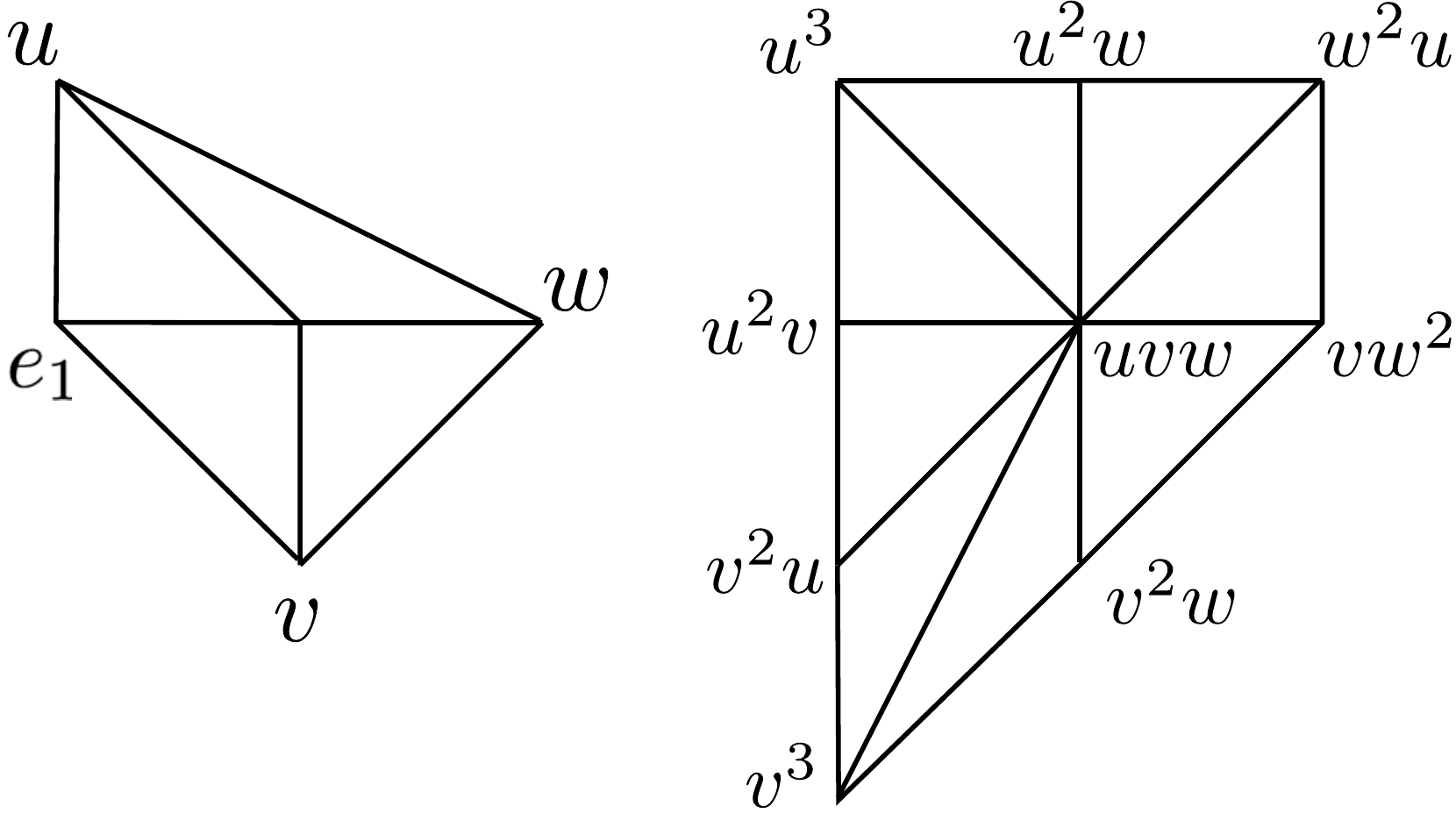}
\end{minipage}%
\begin{minipage}{.44\textwidth}
{\footnotesize
  \begin{tabular}{|c|c|}\hline
Section & Line Bundle\\ \hline
$u$ & $\mathcal{O}(H-E_1+\cS_9+K_B)$ \\ \hline
$v$ & $\mathcal{O}(H-E_1+\cS_9-\cS_7)$\\ \hline
$w$ & $\mathcal{O}(H)$\\ \hline
$e_1$ & $\mathcal{O}(E_1)$\\ \hline
\end{tabular}}
\end{minipage}
\caption{\label{fig:F3_toric} The polytope $F_{3}$ and its dual. The table on the right provides the line bundle classes for the coordinates in $\mathbb{P}_{F_{3}}$.}
\end{figure}

 The F-theory fourfold can be described as a toric hypersurface fibration based on the toric ambient space $\mathbb{P}_{F_3}=dP_1$ as shown in Fig. \ref{fig:F3_toric}.
The hypersurface equation is $p_{F_3}=0$ with
\beq \label{eq:PF3}
	p_{F_3}=s_1 u^3e_1^2 + s_2 u^2 ve_1^2  + s_3 u v^2 e_1^2  +  s_4 v^3e_1^2 +
  s_5 u^2 w e_1
+  s_6 u v w e_1 +  s_7 v^2 w e_1 + s_8 u w^2 + s_9 v w^2 \, .
\eeq
The line bundles of the $s_i$ are fixed by choosing two arbitrary classes, in this case $\cS_7=[s_7]$ and $\cS_9=[s_9]$, and by requiring that all the monomials of \eqref{eq:PF3} are sections of the same line bundle $\mathcal{O}(3H-E_1+2\cS_9-\cS_7)$. After mapping $p_{F_3}$ to the Weierstrass form we obtain $f$, $g$ and $\Delta$, which can be taken from~\cite{Klevers:2014bqa} and are also reported in Appendix B of \cite{MayorgaPena:2017eda}.

The fourfold described by \eqref{eq:PF3} has two sections of the elliptic fibration: the (birationally equivalent) Weierstrass model  zero section $S_0$, and an extra section $S_1$. This gives a massless \U1 gauge symmetry in the low dimensional effective theory. 

There are no codimension-one singularities. At codimension-two one finds three $I_2$ fibers corresponding to states charged under the $\U{1}$ symmetry. The loci for the corresponding states are given in Table~\ref{tab:F3_matter}. 
\begin{table}[ht!]
\begin{center}
\footnotesize
\renewcommand{\arraystretch}{1.4}
\begin{tabular}{|c|c|}
\hline
 Representation&  Locus  \\ \hline
${\bf 1}_{3}$ &  $V(I_{(3)}):=\{s_8 = s_9 = 0\}$   \\ \hline

${\bf 1}_{2}$ &  $
 V(I_{(2)}):=
 \{s_4 s_8^3- s_3 s_8^2 s_9 + s_2 s_8 s_9^2 -s_1 s_9^3
 =s_7 s_8^2  + s_5 s_9^2\!-\! s_6 s_8 s_9 =0\}\backslash\ V(I_{(3)})
 $
 \\ \hline
${\bf 1}_{1}$ & $\begin{array}{c}
V(I_{(1)}):=\{y_{1}=f z_1^4+3 x_1^2=0\}\backslash\ (V(I_{(2)})\cup V(I_{(3)}))
\end{array}$ \\ \hline
\end{tabular}
\caption{\label{tab:F3_matter} The loci for the charged matter representations under the  $\U1$ symmetry. The charges are written as subscripts. The locus for the state ${\bf 1}_{1}$ is given in terms of the sections $x_1$, $y_1$ and $z_1$ discussed in Appendix~\ref{app:Charge3II}.} 
\end{center}
\end{table}

Let us now discuss the type IIB limit of this model. As pointed out already in Section~\ref{Sec:SenLim}, from the facets of the dual polytope we can deduce four types of $\epsilon$-scalings leading to consistent weak coupling limits. Out of these four, only two are inequivalent from the point of view of the brane setups one realizes in type IIB, both of these lead to the same gauge gauge group and matter spectrum of the parent F-theory model. 

\subsubsection*{Sen limit to charge 3 (I) model}
We start with the limit studied in \cite{MayorgaPena:2017eda}. We set the following $\epsilon$ scalings for the sections $s_i$:
\beq
 s_1 \rightarrow \epsilon^1 s_1, \quad s_5 \rightarrow \epsilon^1 s_5, \quad s_8 \rightarrow \epsilon^1 s_8, \quad s_i \rightarrow \epsilon^0 s_i \quad {\rm (}i\neq 1,5,8{\rm )}\:.
\eeq
In the limit $\epsilon\rightarrow 0$ the discriminant of the elliptic fibration factorizes as $\Delta \sim b_2 \Delta_E$.
The location of the O7-plane is at $b_2=0$ with
$$
b_2=\frac{s_6^2}{4} - s_2s_9\:.
$$
The D7-brane locus is $\Delta_E=0$ with 
\begin{align}
\begin{split}
\Delta_E=-\frac14 s_9&\,(s_2^2 s_8^2 + 
   s_2 (-s_5 s_6 s_8 + s_5^2 s_9 - 2 s_1 s_8 s_9) + 
   s_1 (s_6^2 s_8 - s_5 s_6 s_9 + s_1 s_9^2)) \\
  &\times (-s_3 s_6 s_7 + s_2 s_7^2 + s_3^2 s_9 + 
   s_4 (s_6^2-4s_2 s_9)) \,.
\end{split}
\end{align}
We then find the charge 3 (I) model that we constructed from scratch in type IIB, with the identifications $(s_L,s_6,s_R)=(s_9,s_6,s_2)$, $(p_1,p_2)=(0,1)$, $(q_1,q_2,q_3)=(s_1,s_5,s_8)$ and $(w_1,w_2,w_3)=(s_7,s_3,s_4)$.

Applying the limit $\epsilon\rightarrow 0$ to the matter loci in Table~\ref{tab:F3_matter}, one sees that the charge 3 locus become the corresponding locus in type IIB, while the charge 1 (charge 2) locus splits into the two charge 1 (charge 2) loci of the type IIB model \cite{MayorgaPena:2017eda}. We will see how this mechanism works explictly in the next example. For the charge 3 (I) model the computations are reported in \cite{MayorgaPena:2017eda} 


\subsubsection*{Sen limit to the charge 3 (II) model}

One can take a different weak coupling limit, by choosing the following scaling
\beq\label{eq:wcch3II}
 s_1 \rightarrow \epsilon^1 s_1, \quad s_2 \rightarrow \epsilon^1 s_2, \quad s_3 \rightarrow \epsilon^1 s_3, \quad s_4 \rightarrow \epsilon^1 s_4, \quad s_i \rightarrow \epsilon^0 s_i \quad {\rm (}i\neq 1,2,3,4{\rm )}\:.
\eeq
After the limit $\epsilon\rightarrow 0$, 
the location of the O7-plane is at $b_2=0$ with
$$
b_2=\frac{s_6^2}{4}-s_5 s_7\:,
$$
while the D7-brane locus is $\Delta_E=0$ with 
\begin{align}
\begin{split}
\Delta_E=-\frac14 &(s_5 s_9^2 -s_9 s_6 s_8 + s_7 s_8^2) \\
&\times(s_4^2 s_5^3 - 
  s_4 (s_3 s_5^2 s_6 + s_1 s_6 (s_6^2 - 3 s_5 s_7) + s_2 s_5 (-s_6^2 + 2 s_5 s_7)) \\
 &\phantom{\times(} +
   s_7 (s_3^2 s_5^2 + s_7 (s_2^2 s_5 - s_1 s_2 s_6 + s_1^2 s_7) + 
     s_3 (-s_2 s_5 s_6 + s_1 (s_6^2 - 2 s_5 s_7)))) \,.
\end{split}
\end{align}
We then find the type IIB charge 3 (II) model, with the identifications $(s_L,s_6,s_R)=(s_5,s_6,s_7)$, $(p_1,p_2)=(s_9,s_8)$, $(r_1,r_2,r_3,r_4)=(s_4,s_3,s_2,s_1)$. The compete match of codimension two loci between the F-theoretic and the type IIB model is summarized in Appendix~\ref{app:Charge3II}. In this case the intersections $D7_1 \cap D7_2$ and $D7_2 \cap D7_2'$ recombine away from weak coupling. 

\subsection{Charge 4 type IIB model from F-theory}

The first \U1 F-theory model with massless matter charged up to $Q_{\rm max}=4$ was derived in \cite{Raghuram:2017qut}. It can be viewed as a specialized version of the torus hypersurface equation ($\tilde{p}_{F_1}=0$) in the toric ambient space $\mathbb{P}_{F_1}$ \cite{Klevers:2014bqa}. The polynomial equation is given as follows
\begin{equation}
\label{polynomial}
\tilde{p}_{F_1^\prime}=u(s_1u^2+s_2uv+s_3v^2+s_5uw+s_6vw+s_8w^2)+(a_1v+b_1w)(d_0v^2+d_1vw+d_2w^2).
\end{equation}
The divisor classes of the coordinates in $\mathbb{P}_{F_1}$ as well as those for the sections $s_i$, $a_i$ and $d_i$ are given in Table~\ref{table_bundle}. The model is therefore described in terms of three base divisors: $[a_1]$, $[b_1]$ and $[\hat{z}]$ where 
\beq\label{eq:zz}
\hat z=(s_2 b_1-s_5 a_1) \alpha^2+\frac{1}{a_1}(s_6 a_1-2 s_3 b_1)\alpha \beta-\frac{1}{a_1}(d_1 a_1-2 d_0 b_1)\beta^2\,
\eeq
with 
\begin{equation}\label{eq.alphabeta}
\alpha=d_2 a_1^2-d_1 a_1 b_1+d_0 b_1^2\, \quad\mbox{and}\qquad \beta=s_8 a_1^2-s_6 a_1 b_1+s_3 b_1^2\,,
\end{equation}
is the coordinate of the extra section $S_1$ of the elliptic fibration, responsible for the massless~\U1.
\begin{table}[h]
	$$\renewcommand{\arraystretch}{1.4}
	\begin{array}{|c|c|}
	\hline 
	\text{\small{ Section}} & \text{\small{Line Bundle}} \\ 
	\hline 
	a_1 & \mathcal{O}([a_1]) \\ 
	\hline 
	b_1 & \mathcal{O}([b_1]) \\ 
	\hline 
	d_0 & \mathcal{O}([\hat{z}]-2\bar{K}_B-[a_1]-3[b_1]) \\ 
	\hline 
	d_1 & \mathcal{O}([\hat{z}]-2\bar{K}_B-2[a_1]-2[b_1])\\ 
	\hline 
	d_2 & \mathcal{O}([\hat{z}]-2\bar{K}_B-3[a_1]-[b_1]) \\ 
	\hline 
	s_1 & \mathcal{O}(7\bar{K}_B-2[\hat{z}]+3[a_1]+3[b_1]) \\ 
	\hline 
	s_2 & \mathcal{O}(4\bar{K}_B-[\hat{z}]+2[a_1]+[b_1]) \\ 
	\hline 
	s_3 & \mathcal{O}(\bar{K}_B+[a_1]-[b_1]) \\ 
	\hline 
	s_5 & \mathcal{O}(4\bar{K}_B-[\hat{z}]+[a_1]+2[b_1]) \\ 
	\hline 
	s_6 & \mathcal{O}(\bar{K}_B) \\ 
	\hline 
	s_8 & \mathcal{O}(\bar{K}_B-[a_1]+[b_1]) \\ 
	\hline 
	\end{array}
	\quad	\begin{array}{|c|c|}
	\hline 
	\text{\small{Section}} & \text{\small{Line Bundle}} \\ 
	\hline 
	u & \mathcal{O}(H-[a_1]+K_B) \\ 
	\hline 
	v & \mathcal{O}(H+[b_1]-[a_1]) \\ 
	\hline 
	w & \mathcal{O}(H) \\ 
	\hline
	\end{array}
	$$
	\caption{Left: The base sections and their corresponding line bundle classes. Right: The  classes for the sections in the toric ambient space $\mathbb{P}_{F_1}$. }
	\label{table_bundle}
\end{table}

The matter spectrum is given in Table \ref{tab:Charge4_matter}, we stick to the notation of \cite{Raghuram:2017qut} and refer to it for further details. For our purposes it suffices to specify the expression
\begin{align}\label{eq:tt}
\begin{split}
t=&s_1 \alpha^3+(s_2 s_6 b_1-s_2 s_8 a_1-s_5 s_3 b_1)\alpha^2+(2 s_3 s_8-s_6^2)\alpha \beta\\
&-\beta^2(s_8 d_0-s_6 d_1+s_3 d_2)+a_1 s_6(d_0 s_8 b_1-s_3 d_2 b_1+s_6 d_2 a_1-a_1 d_1 s_8)\beta\,.
\end{split}
\end{align}


\begin{table}[ht!]
\begin{center}
\footnotesize
\renewcommand{\arraystretch}{1.4}
\begin{tabular}{|c|c|}
\hline
 Representation&  Locus  \\ \hline
 ${\bf 1}_{4}$ &  $V(I_{(4)}):=\{a_1 = b_1 = 0\}$   \\ \hline
${\bf 1}_{3}$ &  $V(I_{(3)}):=\{\alpha =\beta = 0\}\backslash\ V(I_{(4)})$   \\ \hline
${\bf 1}_{2}$ &  $
 V(I_{(2)}):=
 \{t=\hat{z}=0\}\backslash\ V(I_{(4)})\cup V(I_{(3)})
 $
 \\ \hline
${\bf 1}_{1}$ & $\begin{array}{c}
V(I_{(1)}):=\{\hat y=f \hat z^4+3 \hat x^2=0\}\backslash\ V(I_{(4)})\cup V(I_{(3)})\cup (V(I_{(2)})
\end{array}$ \\ \hline
\end{tabular}
\caption{\label{tab:Charge4_matter} The loci for the charged matter representations under the  $\U1$ symmetry for the charge 4 model. The charges are written as subscripts. The locus for the state ${\bf 1}_{1}$ is given in terms of the sections $\hat x$, $\hat y$ and $\hat z$ that are the coordinates for the additional section discussed in Appendix~\ref{app:Charge4}.}
\end{center}
\end{table}

Let us now consider the following weak coupling limit 
\begin{equation}
a_1\rightarrow \epsilon^1 a_1\,, \quad  b_1\rightarrow \epsilon^1 b_1\,,  \quad  d_i\rightarrow \epsilon^0 d_i\,, \quad  s_i\rightarrow \epsilon^0 s_i\:.
\end{equation}
We then have
\beq
b_2=\frac{s_6^2}{4}-s_3 s_8 \qquad\mbox{and}\qquad \Delta_E=-\frac{1}{4} \Delta_1\cdot\Delta_2\cdot \Delta_3 \:,
\eeq
with the irreducible components $\Delta_i$ given by
\begin{eqnarray}
\label{delta_remain}
\Delta_1&\equiv& b_1^2 s_3 - a_1 b_1 s_6 + a_1^2 s_8,\nonumber \\
\Delta_2&\equiv& s_3 s_5^2 - s_2 s_5 s_6 + s_1 s_6^2 + s_2^2 s_8 - 4 s_1 s_3 s_8,\\
\Delta_3&\equiv& d_2^2 s_3^2 - d_1 d_2 s_3 s_6 + d_0 d_2 s_6^2 + d_1^2 s_3 s_8 - 2 d_0 d_2 s_3 s_8 -d_0 d_1 s_6 s_8 + d_0^2 s_8^2.\nonumber
\end{eqnarray}
The first locus corresponds to a pair of massive \U1 brane/image-brane of the form \eqref{EqD7p1p2}, the second is an invariant brane of the type \eqref{invD7ord1} and the third is a pair of brane/image-brane of the type \eqref{EqD7q1q2}.
Hence we find the configuration predicted in Section~\ref{sec:odd}, with the identifications $(s_L,s_6,s_R)=(s_3,s_6,s_8)$, $(p_1,p_2)=(b_1,a_1)$, $(w_1,w_2,w_3)=(s_2,s_5,s_1)$ and $(q_1,q_2,q_3)=(d_2,d_1,d_0)$ The match of the charged loci with the   type~IIB ones is provided in Appendix \ref{app:Charge4}.

\subsection{Charge 4 in F-theory from a $\mathbb{Z}_3$ model through type~IIB}\label{Sec:Charge4FromTth}

In \cite{Raghuram:2017qut}, the author found the charge 4 model by Higgsing a non-generic $\U1 \times \U1$ model \cite{Cvetic:2015ioa}. In this section we show that this model can be obtained also from the model in \cite{Klevers:2014bqa}, i.e. a fourfold with  $\mathbb{Z}_3$ discrete symmetry. We will use the weak coupling limit to achieve this result.

The genus one fibration describing the $\mathbb{Z}_3$ fiber in $\mathbb{P}_{F_1}=\mathbb{P}^2$ is given by the equation
\begin{equation}\label{Z3elliptFib}
p_{F_{1}} = s_1 u^3 + s_2 u^2 v+ s_3 u v^2 + s_4 v^3 + s_5 u^2 w + s_6 u v w + s_7 v^2 w + s_8 u w^2+s_9 v w^2 +s_{10} w^3\,.
\end{equation}

The Sen limit of the $\mathbb{Z}_3$ was studied in \cite{MayorgaPena:2017eda}, with the scaling
 \beq
 s_4\rightarrow \epsilon s_4\,,\quad s_7\rightarrow \epsilon s_7\,, \quad s_9\rightarrow \epsilon s_9\,, \quad s_{10}\rightarrow \epsilon s_{10}\,. 
 \eeq 
The type~IIB double cover CY $X$ is given by the equation $\xi^2=\frac{s_6^2}{4}-s_Ls_R$. with $s_L=s_3$ and $s_R=s_8$. The weak coupling discriminant is
\begin{align}\label{Z3DeltaE}
\begin{split}
\Delta_E=-\frac14 &\, \left(s_2^2 s_8 -s_2 s_5 s_6 + s_5^2s_3 + 4 s_1 \left(\frac{s_6^2}{4} - s_3  s_8\right)\right)\\
    &\times \left[ s_{10}^2 s_3^3 - 
 s_{10} (s_4 (s_6^3 - 3 s_3 s_6 s_8) + 
    s_3 (-s_6^2 s_7 + 2 s_3 s_7 s_8 + s_3 s_6 s_9)) \right.\\
    &\,\,\,\,+ 
 \left. s_8 (s_4^2 s_8^2 + s_4 (-s_6 s_7 s_8 + s_6^2 s_9 - 2 s_3 s_8 s_9) + 
    s_3 (s_7^2 s_8 - s_6 s_7 s_9 + s_3 s_9^2)) \right]\,.
\end{split}\nonumber
\end{align}
The D7-brane configuration is given by a massive \U1 D7-brane and an invariant brane. The \U1 brane wraps a divisor defined by the equation \eqref{DB2inquot3}  in the quotient $B$ with $(r_1,r_2,r_3,r_4)=(s_{10},s_9,s_7,s_4)$, while the invariant brane wraps a divisor described by equation \eqref{invD7ord1} with $(w_1,w_2,w_3)=(s_5,s_2,s_1)$. 

At weak coupling, one may wonder which restriction on the sections $r_i$ makes the \U1 brane locus \eqref{DB2inquot3} factorize into two massive branes, one with equation \eqref{EqD7p1p2} ($(p_1,p_2)=(b_1,a_1)$) and one with equation \eqref{EqD7q1q2} ($(q_1,q_2,q_3)=(d_2,d_1,d_0)$), i.e. one needs to find the expressions of $r_{1,...,4}$ in terms of $p_{1,2}$ and $q_{1,2,3}$ that solve the equation
\begin{equation}
   \vec{r}\cdot \mathrm{A}_3\vec{r} \,\,=\,\, \left(   \vec{p}\cdot \mathrm{A}_1\vec{p} \right)\,\left(   \vec{q}\cdot \mathrm{A}_2\vec{q}\: \right) \:.
\end{equation}
This problem has a definite answer, that in the F-theory model notation is
\begin{equation}\label{RestrictionZ3toQ4}
s_{10} = b_1 d_2\,,\quad  s_9= b_1 d_1 + a_1 d_2\,, \quad s_7 = b_1 d_0 + a_1 d_1\,, \quad  s_4=a_1 d_0\,.
\end{equation}
The resulting D7-brane configuration is the one realizing the charge 4~(I) type~IIB model. This result is just obtained in the perturbative limit of the $\mathbb{Z}_3$ model. If one applies the same restriction \eqref{RestrictionZ3toQ4} to the equation \eqref{Z3elliptFib}, one gets the equation
\begin{equation}
p_{F_{1}}^{\prime}= s_1 u^3 + s_2 u^2 v+ s_3 u v^2 + s_5 u^2 w + s_6 u v w + s_8 u w^2+(a_1 v+b_1 w)(d_0 v^2+d_1 v w+d_2 w^2)\,.
\end{equation}
that is exactly the F-theory charge 4 model found in \cite{Raghuram:2017qut}.

This computation shows how the weak coupling limit can help to construct explicit complicated models in a simple way. We leave for the future the application of  this method to higher charge models.

\section{Charge 5 and 6 models in type~IIB}\label{Sec:charge56}

As we have seen in Section~\ref{sec:minimal}, there are not many possibilities to construct a model with maximal charge 5 or 6 when $h^{1,1}_{-,{\rm eff}}=1$. There is only one charge 6 model and in principle two charge 5 ones. Actually only the charge 5 (II) model can be constructed in full generality. The charge 5 (I) model requires a brane/image-brane pair wrapping the locus \eqref{DB2inquot3} with zero intersection away from the orientifold locus; this is not as easy to realize as when they wrap a locus like \eqref{DB2inquot3}; only for specific base manifolds this may be realized. For this reason we will explicitly describe only the charge 5 (II) model and the charge 6 one.

\subsection{Charge 5 (II) model in type~IIB}

In this case we have the following homology relation among the odd cycles:
\begin{equation}
  \mathcal{D}_{D7_2} -  \mathcal{D}_{D7_2'} = -4( \mathcal{D}_{D7_1} -  \mathcal{D}_{D7_1'})\:,
\end{equation}
which is consitent with the choice $a=1$, $b=4$ for the generator of the massless \U1. Note that the intersections among  $\mathcal{D}_{D7_1}$  and $ \mathcal{D}_{D7_1'}$ are required to vanish away from the orientifold locus. This is possible whenever we specialize the vector $\vec{p}$ in \eqref{DBp1p2} to be of the form $(p_1,p_2)=(1,0)$ such that the brane locus in the weak coupling discriminant reads\footnote{Here one could as well take the divisor $P_{D7_1}=s_R$ for the choice $(p_1,p_2)=(0,1)$.}
$P_{D7_1}=0$ with 
\begin{equation}
P_{D7_1}=s_L\,.
\end{equation}

The locus for the brane $\mathcal{D}_{D7_2}$ and its image must be of order four in $s_L$, $s_6$ and $s_R$, and it can be written as, 
\begin{equation}\label{Div72Ch5}
P_{D7_2}= \begin{pmatrix}
 t_1 & t_2 & t_3 & t_4 & t_5 \\
\end{pmatrix} \cdot \mathrm{A}_4 \cdot   \begin{pmatrix}
 t_1 \\ t_2 \\ t_3 \\ t_4 \\ t_5 \\
\end{pmatrix} \,.
\end{equation}
The matrix $\mathrm{A}_4$ can be obtained as the non-trivial $5 \times 5$ block inside $\tilde {\rm M}^{\otimes 4}$ and it is given in~\eqref{eq:MF4}. Equation \eqref{Div72Ch5} finally reads 
\begin{align}\label{Div72Ch52}
\begin{split}
P_{D7_2}=&s_L^4 t_1^2 + s_L^3 (-s_6 t_1 t_2 + s_R (t_2^2 - 2 t_1 t_3)) + 
 t_5 (s_6^4 t_1 - s_6^3 s_R t_2 + s_6^2 s_R^2 t_3 - s_6 s_R^3 t_4 + s_R^4 t_5)\\
 &+ 
 s_L^2 (s_6^2 t_1 t_3 + s_6 s_R (-t_2 t_3 + 3 t_1 t_4) + 
    s_R^2 (t_3^2 - 2 t_2 t_4 + 2 t_1 t_5))\\
    & + 
 s_L (-s_6^3 t_1 t_4 + s_6^2 s_R (t_2 t_4 - 4 t_1 t_5) + 
    s_6 s_R^2 (-t_3 t_4 + 3 t_2 t_5) + s_R^3 (t_4^2 - 2 t_3 t_5))\,,
    \end{split}\nonumber
\end{align}
with the $D7_2$ locus given by $P_{D7_2}=0$. 
The invariant brane locus is given by $P_{D7_{\rm inv}}=0$, with the polynomial $P_{D7_{\rm inv}}$ given in~\eqref{DB2inquot3}. The  D7-brane configuration is therefore described by $\Delta_E = P_{D7_1}  P_{D7_2}  P_{D7_{\rm inv}}$. Note that from the weak coupling perspective we require  11  sections of line bundles on $B$ in order to describe the charge five model: $s_{6,L,R}$, $w_{1,2,3}$, $t_{1,...,5}$. These must become the sections $\mathsf{s}_\kappa$ defining the F-theory model.

The charges of the fields for the type~IIB perspective are given in Table~\ref{tab:Ch5IIB}.
\begin{table}[h!]
\begin{center}
\renewcommand{\arraystretch}{1.4}
{\footnotesize
\begin{tabular}{|c|c|c|c|c|c|c|}\cline{2-7}
\multicolumn{1}{c|}{} & $D7_1\cap D7_2$ &  $D7_1\cap D7_2'$ & $D7_1\cap D7_1'$ & $D7_2\cap D7_2'$  & $D7_1\cap D7_{\rm inv}$  & $D7_2\cap D7_{\rm inv}$  \\ \hline
$(Q_1,Q_2)$ & $(1,-1)$ & $(1,1)$ & $(2,0)$ & $(0,2)$& $(1,0)$  & $(0,1)$   \\ \hline\hline
$Q$ & $3$ & $5$ & $ - $ & $2$ & $4$ & $1$ \\ \hline
\end{tabular}
}
\caption{\label{tab:Ch5IIB} \U1 charges for the charge 5 (II) model. The state at $D7_1\cap D7_1'$ is absent.} 
\end{center}
\end{table}

As a final remark, notice that in this model we have the peculiarity that all the matter loci exhibit different charges under the massless \U1 therefore in the F-theory uplift we expect no recombination phenomena occurring for the codimension two loci.

\subsection{Charge 6 model in type~IIB}\label{Sec:Charge6Model}

The charge six model at weak coupling is obtained 
with $a=3$, $b=2$ and $\kappa=1$. We will have the following relation among the odd divisors:
 \begin{equation}
 2( \mathcal{D}_{D7_2} -  \mathcal{D}_{D7_2'}) = -3( \mathcal{D}_{D7_1} -  \mathcal{D}_{D7_1'})\,.
\end{equation}
The D7-brane configuration is given by $\Delta_E = P_{D7_1}  P_{D7_2}  P_{D7_{\rm inv}}$ with $P_{D7_1}$ given by \eqref{DB2inquot}, i.e.
\begin{equation}\label{EqD7deg2}
 P_{D7_1} = \left( s_Lq_1-s_Rq_3 \right)^2  - \left( s_Lq_2-s_6q_3 \right)\left( s_6q_1-s_Rq_2 \right)\:,
 \end{equation}
and $ P_{D7_2}=0$ as in \eqref{DB2inquot3}, i.e.
\begin{align}
\begin{split}\label{EqD7deg3}
 P_{D7_2} =&\, r_1^2 s_L^3 - r_1 s_6 (r_4 s_6^2 + s_L (-r_3 s_6 + r_2 s_L)) + 
 r_1 s_L (3 r_4 s_6 - 2 r_3 s_L) s_R \\
 &+ 
 s_R (r_2^2 s_L^2 + s_R (-r_3 r_4 s_6 + r_3^2 s_L + r_4^2 s_R) + 
    r_2 (r_4 s_6^2 - r_3 s_6 s_L - 2 r_4 s_L s_R))\:.
    \end{split}
 \end{align}
The invariant brane is at $P_{D7_{\rm inv}}=0$ with  $P_{D7_{\rm inv}}$ given in terms of the base sections $(w_1,w_2,w_3)$ according to \eqref{invD7ord1}. For completeness we summarize the charges for the fields in Table~\ref{tab:Ch6IIB}.
\begin{table}[h!]
\begin{center}
\renewcommand{\arraystretch}{1.4}
{\footnotesize
\begin{tabular}{|c|c|c|c|c|c|c|}\cline{2-7}
\multicolumn{1}{c|}{} & $D7_1\cap D7_2$ &  $D7_1\cap D7_2'$ & $D7_1\cap D7_1'$ & $D7_2\cap D7_2'$  & $D7_1\cap D7_{\rm inv}$  & $D7_2\cap D7_{\rm inv}$  \\ \hline
$(Q_1,Q_2)$ & $(1,-1)$ & $(1,1)$ & $(2,0)$ & $(0,2)$& $(1,0)$  & $(0,1)$   \\ \hline\hline
$Q$ & $1$ & $5$ & $ 6 $ & $4$ & $3$ & $2$ \\ \hline
\end{tabular}
}
\caption{\label{tab:Ch6IIB} \U1 charges for the charge 6 model.} 
\end{center}
\end{table}

Note that all different intersections have different charges under the massless \U1 symmetry and therefore we expect no recombination for these loci in the F-theory lift. In this case we need 13 base sections in order to fully describe a charge 6 model at weak coupling: $s_{6,L,R}$, $q_{1,2,3}$, $r_{1,...,4}$ and $w_{1,2,3}$. We claim that these will be the $\mathsf{s}_\kappa$ sections describing the charge 6 F-theory model.

\subsection{Explicit examples with charge 5 and 6}

We now proceed to construct
explicit 6 dimensional charge 5 and charge 6 models in perturbative type~IIB. We choose the two-dimensional base manifold to be $B=\mathbb{P}^2$, with homogeneous coordinates $[x_1,x_2,x_3]$. The double cover CY two-fold is described by $\xi^2=b_2$, where $b_2$ is a section of $\mathcal{O}(6)$ ($\L=\bar{K}_{\mathbb{P}^2}=\mathcal{O}(3)$). Hence the O7-plane is in the class $[O7]=3H$.
We choose also the line bundle $\L_L=\bar{K}_{\mathbb{P}^2}$, so that we have $\L_R=\bar{K}_{\mathbb{P}^2}$ as well. 
This means that the classes of $s_6,s_L,s_R$ are 
\begin{equation}
[s_6]=[s_L]=[s_R]=3H,
\end{equation} 
i.e. they are homogeneous polynomials of degree $3$ in the homogeneous coordinates $x_i$.\footnote{This choice is the only one that allows to satisfy the D7-brane tadpole, with the chosen base manifold.} These polynomials can be taken independent of each other.\footnote{The maximal number of independent polynomials of degree $3$ in three variables is equal to $10$.} 

The CY $X$ is a $K3$ surface and the orientifold involution is one of the Nikulin involution with $k=0$. 
In the present example, the fixed point locus $\xi=0$ is connected and has genus
\begin{equation}
 g = \frac12(2-\chi)=\frac12\left(2 - \int_{O7} c_1(O7) \right)= \frac12\left(2 + \int_{X} 9H^2 \right) =10\:.
\end{equation}
The only involution compatible with $g=10$ and $k=0$, according to Nikulin, is $(r,\mathrm{a},\delta)=(1,1,1)$.
Accordingly we have only one even two-cycle (as we expect from $b_2(\mathbb{P}^2)=1$). Since $b_2(K3)=22$, we have several odd divisor classes. Choosing $b_2=\frac{s_6^2}{4}-s_Ls_R$, we have restricted the complex structure in a way to make algebraic two image two-cycles in different homology classes (without generating singularities in $K3$).

\subsubsection*{Charge 5 model}

Let us consider first the model with charge 5. In order to work out the homology classes of the brane stacks $D7_1/D7_1'$, $D7_2/D7_2'$ and $D7_{\rm inv}$ we notice that the choice of $\L_L$ implies that the divisor class for the sections $t_i$ in~\eqref{Div72Ch5} are all equal to each other, $[t_1]=...=[t_5]=\beta H$ with $\beta$ a positive integer number. As regard the invariant brane locus  \eqref{invD7ord1} we have $[w_1]=[w_2]\equiv \nu H$ and $[w_3]=(2\nu-3)H$. Effectiveness of these divisor classes implies that  $\nu$ is an integer greater or equal than two. The classes of the weak coupling discriminant loci are therefore given by: 
\begin{eqnarray}
\left[P_{D7_1}\right] &=&3H\,, \\
 \left[ P_{D7_2} \right] &=&	12H + 2 [t_1] = 2(6 + \beta)H\,,	\\
\left[ P_{D7_{\rm inv}} \right] &=& 6H +[w_3] = (3+2\nu)H\,.		
\end{eqnarray}
D7-tadpole cancellation implies that 
\begin{equation}
\left[P_{D7_1}\right] + \left[P_{D7_2}\right] + \left[ P_{D7_{\rm inv}} \right] = 8 [O7]\,,
\end{equation}
conversely
\begin{equation}
 3+ 2(6+\beta)+ (3+2\nu) = 8 \cdot 3 \qquad\mbox{i.e.}\qquad \beta+\nu=3\:.
\end{equation}
and therefore the only two possibilities for $(\beta,\nu)$ are $(1,2)$ and $(0,3)$. 
The matter multiplicities for these states are given in Table~\ref{tab:HyperMultch5}.
\begin{table}[h!]
\begin{center}
\renewcommand{\arraystretch}{1.4}
{\footnotesize
\begin{tabular}{|c|c|c|c|c|}\cline{2-4}
\multicolumn{1}{c|}{} & $(\beta,\nu)$ & $(1,2)$ & $(0,3)$ \\ \hline
$\mathbf{1}_1$ & $36 + 3 \beta$ & 39 & 36 \\
$\mathbf{1}_2$ & $ 36 + 6 \beta + 24 \nu + 4 \beta \nu$ & 98 & 108  \\
 $\mathbf{1}_3$ & $ 9 + 6 \nu$ & 21 & 27 \\
 $\mathbf{1}_4$  & $54 + 9 \beta + \beta^2$& 64 & 54  \\
  $\mathbf{1}_5$  & $3 \beta$& 3 & 0 \\ \hline
\end{tabular}
}
\caption{\label{tab:HyperMultch5} Hypermultiplet multiplicities for the charged matter in the two explicit charge 5 models constructed in the double cover $K3$ over $\mathbb{P}^2$.} 
\end{center}
\end{table}

The type IIB analogous of the Nèron-Tate height pairing that enters the anomaly cancellation in 6D (see Appendix~\ref{App:AnomalyCancel}) is given by
\beq
-\mathsf{b}=[3+32(6+\beta)]H\,.
\eeq

\subsubsection*{Charge 6 model}

We now choose a D7-brane configuration realizing $Q_{\rm max}=6$.
We have 
three stacks of branes $D7_1/D7_1'$, $D7_2/D7_2'$ and $D7_{\rm inv}$, as described in Section~\ref{Sec:Charge6Model}. We notice that the choice we made on $\L_L$ also implies that 
$[q_1]=[q_2]=[q_3]\equiv \lambda H$, $[r_1]=[r_2]=[r_3]=[r_4]\equiv \rho H$, $[w_1]=[w_2]\equiv \nu H$ and $[w_3]=(2\nu-3)H$, with $\lambda,\rho,\nu\in \mathbb{N}$ and $\nu\geq 2$. Hence, considering the equations \eqref{EqD7deg2}, \eqref{EqD7deg3} and \eqref{invD7ord1} we obtain
\begin{eqnarray}
\left[P_{D7_1}\right] &=&	6H + 2 [q_1] = 2 (3+\lambda)H \:,\\
 \left[ P_{D7_2} \right] &=&	9H + 2 [r_1] = (9 + 2\rho)H	\:,\\
\left[ P_{D7_{\rm inv}} \right] &=& 6H +[w_3] = (3+2\nu)H	\:.	
\end{eqnarray}
D7-tadpole cancellation implies that 
\begin{equation}
\left[P_{D7_1}\right] + \left[P_{D7_2}\right] + \left[ P_{D7_{\rm inv}} \right] = 8 [O7]
\end{equation}
that means
\begin{equation}
  2(3+\lambda) + (9+2\rho) + (3+2\nu) = 8 \cdot 3 \qquad\mbox{i.e.}\qquad \lambda+\rho+\nu=3\:.
\end{equation}
Since $\nu\geq 2$, the only possibilities for  $(\lambda,\rho,\nu)$ are $(1,0,2)$, $(0,1,2)$ or $(0,0,3)$.
Each of these choices produces an explicit smooth model in type~IIB with one massless \U1 and matter with charges $1,2,3,4,5,6$. The matter multiplicities for these states are given in Table~\ref{tab:HyperMult}.
\begin{table}[h!]
\begin{center}
\renewcommand{\arraystretch}{1.4}
{\footnotesize
\begin{tabular}{|c|c|c|c|c|c|}\cline{2-5}
\multicolumn{1}{c|}{} & $(\lambda,\rho,\nu)$ & $(1,0,2)$ & $(0,1,2)$ & $(0,0,3)$ \\ \hline
$\mathbf{1}_1$ & $54 + 9 \lambda + 6 \rho + 2 \lambda \rho$ & 63 & 60 & 54 \\
$\mathbf{1}_2$ & $ 27 + 18 \nu + 6 \rho + 4 \nu \rho$ & 63 & 77 & 81 \\
 $\mathbf{1}_3$ & $18 + 6 \lambda + 12 \nu + 4 \lambda \nu$ & 56 & 42 & 54 \\
 $\mathbf{1}_4$  & $27 + 6 \rho + \rho^2$& 27 & 34 & 27 \\
  $\mathbf{1}_5$  & $ 9 \lambda + 6 \rho + 2 \lambda \rho$& 9 & 6 & 0\\
  $\mathbf{1}_6$  & $9 + 3 \lambda + \lambda^2$ & 13 & 9 & 9 \\ \hline
\end{tabular}
}
\caption{\label{tab:HyperMult} Hypermultiplet multiplicities for the charged matter in the three explicit charge 6 models.} 
\end{center}
\end{table}

The type IIB analogous of the Nèron-Tate height pairing (see Appendix~\ref{App:AnomalyCancel}) for this model is given by 
\beq
-\mathsf{b}=[8(3+\lambda)+9(3+2\rho)]H\,.
\eeq
Recall once again that all anomalies are canceled for these models.

\subsection{Models with incomplete spectra beyond charge 6}

In the models we constructed so far, we have
demanded that all charges between 1 and a given maximum charge $Q_{\rm max}$ appear in the spectrum. As we said, since one can have at most six intersections with the D7-brane configuration studied so far, the maximum charge is six. However the bound   $a+b\leq 8$,  derived by the effectiveness of the divisor classes, allows in principle higher charges.

Take for example the case of two $\U1$ branes without an invariant brane. As discussed around equation \eqref{TadpConstrab} this system is bounded by the constraint $a+b\leq 8$ and hence we will have additional options beyond the ones indicated in Section \ref{sec:minimal}: these are going to have only four types of massless charged matter. The possibilities are shown in Table \ref{tab:IncompleteNoInv} where we can see that a model with charge 7 is possible.  
\begin{table}[h!]
\begin{center}
\renewcommand{\arraystretch}{1.4}
{\footnotesize
\begin{tabular}{|ccc|cccc|}
\hline
$a$ & $b$ & $\kappa$ &$D7_1\cap D7_1'$ & $D7_2\cap D7_2'$ & $D7_1\cap D7_2'$ & $D7_1\cap D7_2$ \\ \hline
5 & 1 & 2 & 5 & 1 & 3 & 2 \\
7 & 1 & 2 & 7 & 1 & 4 & 3 \\
5 & 3 & 2 & 5 & 3 & 4 & 1 \\
\hline
\end{tabular}
}
\caption{\label{tab:IncompleteNoInv} $\U1$ charges for the incomplete models with no invariant brane.} 
\end{center}
\end{table}

A similar analysis can be done for the cases in which the invariant brane is present. In that case one would be able to get maximum charge 10. 

\begin{table}[h]
\begin{center}
\renewcommand{\arraystretch}{1.4}
{\footnotesize
\begin{tabular}{|ccc|cccccc|}\hline
$a$& $b$ & $\kappa$ & $D7_1\cap D7_2$ &  $D7_1\cap D7_2'$ & $D7_1\cap D7_1'$ & $D7_2\cap D7_2'$  & $D7_1\cap D7_{\rm inv}$  & $D7_2\cap D7_{\rm inv}$  \\ \hline
3 & 1 & 1 & 2 & 4 & 6 & 2 & 3 & 1 \\
4 & 1 & 1 & 3 & 5 & 8 & 2 & 4 & 1\\
5 & 1 & 1 & 4 & 6 & 10 & 2 & 5 & 1
 \\ \hline
\end{tabular}
}
\caption{\label{tab:Incomplete} \U1 charges for the incomplete models with invariant brane.} 
\end{center}
\end{table}

We find no consistency conditions that would prevent these models in type~IIB, even though the absence of some charge in the spectrum sounds odd in light of F-theory constructions. To realize generic models with $Q_{\rm max}>6$ and with all charges up to $Q_{\rm max}$, one needs to increase the number of (massive \U1) D7-brane stacks.

\section{Conclusions and future directions}\label{Sec:conclusion}

In this paper we have faced the problem of constructing 6D F-theory models with $\U1$ gauge group and matter fields with high charge. These models typically have massless states with integral charges from $1$ to a maximal value $Q_{\rm max}$. 
Our approach was to construct these models in the type~IIB perturbative limit of F-theory. 
If a model with high charge exists in type~IIB, it must have a consistent F-theory lift.

We have worked out a method to easily construct type~IIB models with high charge. We verified that for $Q_{\rm max}=3,4$ these are exactly what one obtains by taking the Sen weak coupling limit of the existing F-theory models with $Q_{\rm max}=3,4$. We then described type~IIB models with $Q_{\rm max}=5,6$. We have built explicit examples, where the base manifold $B$ is $\mathbb{P}^2$. These  are consistent string theory models exhibiting a massless $\U1$ symmetry with massless hypermultiplets with charges up to five and six. This proves therefore that massless states with $\U1$ charges as high as six are part of the 6D string theory landscape. 
Unfortunately, while it is relatively easy to go from an F-theory model to its weak coupling limit, it is not straightforward to lift a type~IIB model to F-theory. 
Nevertheless, the knowledge of what should be its 
Sen limit, can help towards the realization of the corresponding model in F-theory. We plan to approach this issue in the~future.

So far we have explored only models with $Q_{\rm max}\leq 6$. This has been realized by two pairs of \U1 D7 brane/image-brane: one combination of the two \U1 gets a mass by eating a $C_2$ axion, while the orthogonal combination remains massless.
To obtain higher charges one needs to generalize the construction presented in this paper, adding more  D7-branes with massive \U1's  and allowing more $C_2$ axions to be eaten by the massive gauge bosons, i.e. $h^{1,1}_{-,{\rm eff}}>1$ in the notation of Section~\ref{Sec:massiveVSmasslessU1}. Algebraically one can realize it by a more specific form of $b_2$ that admits more than two inequivalent matrix factorizations.
This will lead, without any obstruction, to models with $Q_{\rm max}> 6$. It would be nice to see if there is an upper bound for $Q_{\rm max}$ in type~IIB. In \cite{Raghuram:2018hjn}, it was shown that models with $\U{1}$ symmetries with higher charges can in principle be obtained from models with exotic non-Abelian matter by means of Higgsing. 
Along the same line, it has been shown that $\SU{N}$ models with exotic matter could lead, upon Higgsing to $\U1$ models with charges $Q\leq 21$ in six dimensions. It would be nice to find a similar bound in type~IIB \U1 models (even though with different techniques, as for example three-index antysymmetric states are not realized in perturbative type~IIB).

Another way to obtain models with high \U1 charge is to consider models with gauge symmetry $\U1^n$. As the number of $\U1$'s increases, the number of charged massless fields increases as well. Higgsing models with multiple $\U1$'s could lead to single $\U1$ models with higher charges. For example, the $\U1\times\U1$ model of \cite{Cvetic:2015ioa} has multiplets with charges $(-1,1)$ and $(-2,-2)$ among others. A vev in $(-1,1)$ makes the field $(-2,-2)$ to pick a charge $4$ along the diagonal massless $\U1$ \cite{Raghuram:2017qut}.
This Higgsing can be done in type~IIB as well as in F-theory. Understanding which deformations realize the Higgsing may be easier in type~IIB in some cases. Applying then the same deformation to F-theory models may lead to the desired high charge realizations.

One may apply the same reasoning to models with discrete symmetry. F-theory models exhibiting only discrete symmetries are  described by genus one fibrations.
In this paper we have considered the $\mathbb{Z}_3$ model of \cite{Klevers:2014bqa} and applied the weak coupling limit, obtaining a $\mathbb{Z}_3$ model in type~IIB, realized by a pair of brane/image-brane with massive \U1 and an invariant brane. It was easy to see under which deformation the brane/image-brane system splits into the two brane/image-brane system realizing the $Q_{\rm max}=4$ model. Applying the same deformation to the corresponding $\mathbb{Z}_3$ F-theory model, we were able to straightforwardly obtain the $Q_{\rm max}=4$ F-theory model of \cite{Raghuram:2017qut}. This method can in principle be applied to obtain a charge $Q_{\rm max}>4$ model in F-theory. Models with discrete symmetries $\mathbb{Z}_4$ \cite{Braun:2014qka} and $\mathbb{Z}_{n}$ with $n\leq 5$ \cite{Kimura:2016crs}
are already present in literature.

\section*{Acknowledgments}

We would like to thank A. P. Braun, N. Cabo-Bizet, A. Collinucci, M. Cveti\v{c}, L. Lin, O. Loaiza-Brito, P. Oehlmann and N. Raghuram for inspiring and clarifying discussions. D. M. would like to thank the physics departments at Virginia Tech, the University of Pennsylvania and Northeastern University for hospitality during completion of this work. The work of D.M. is partially supported by CONACyT project N. 258982 and DAIP-Universidad de Guanajuato project N. CIIC 154/2018.
\appendix

\section{Divisors of order 4 in $s_6,s_L,s_R$}\label{App:matrixesn4}
Here we provide the explicit matrix form of $A_4$ and $B_4$:

\begin{equation}\label{eq:MF4}
\renewcommand{\arraystretch}{1.8}
\mathrm{A}_4={\tiny\begin{pmatrix} s_L^4 & -\frac{ s_6 s_L^3}{2} &  -\frac{s_L^2 (-s_6^2 + 2 s_L s_R)}{2} & 
 \frac{s_6 s_L (-s_6^2 + 3 s_L s_R)}{2} & 
  \frac{s_6^4 - 4 s_6^2 s_L s_R + 2 s_L^2 s_R^2}{2}\\
   -\frac12 s_6 s_L^3 &
  s_L^3 s_R &  -\frac{s_6 s_L^2 s_R}{2} & -\frac{s_L s_R (-s_6^2 + 2 s_L s_R)}{2} &
\frac{s_6 s_R (-s_6^2 + 3 s_L s_R)}{2} \\
-\frac{s_L^2 (-s_6^2 + 2 s_L s_R)}{2} & -\frac{s_6 s_L^2 s_R}{2} &
  s_L^2 s_R^2 &  -\frac{s_6 s_L s_R^2}{2} & -\frac{s_R^2 (-s_6^2 + 2 s_L s_R)}{2} \\
  \frac{s_6 s_L (-s_6^2 + 3 s_L s_R)}{2} & -\frac{s_L s_R (-s_6^2 + 2 s_L s_R)}{2} & -\frac{s_6 s_L s_R^2}{2} & s_L s_R^3 & -\frac{s_6 s_R^3}{2} \\
  \frac{s_6^4 - 4 s_6^2 s_L s_R + 2 s_L^2 s_R^2}{2} & \frac{s_6 s_R (-s_6^2 + 3 s_L s_R)}{2} & -\frac{s_R^2 (-s_6^2 + 2 s_L s_R)}{2} & -\frac{s_6 s_R^3}{2} & s_R^4
 \end{pmatrix}}\,,
\end{equation}
and
\begin{equation}
\mathrm{B}_3={\scriptsize\begin{pmatrix} 0 & -s_L^3 & s_6 s_L^2 & -s_6^2 s_L + s_L^2 s_R & s_6^3 - 2 s_6 s_L s_R\\
s_L^3 &  0 & -s_L^2 s_R & s_6 s_L s_R & -s_6^2 s_R + s_L s_R^2\\
-s_6 s_L^2 & s_L^2 s_R & 0 & -s_L s_R^2 & s_6 s_R^2\\
s_6^2 s_L - s_L^2 s_R & -s_6 s_L s_R & s_L s_R^2 & 0 & -s_R^3\\
-s_6^3 + 2 s_6 s_L s_R & s_6^2 s_R - s_L s_R^2 & -s_6 s_R^2 & s_R^3 & 
  0
 \end{pmatrix}}\,.
\end{equation}
Again $\mathrm{A}_4=\mathcal{I}_4\cdot \mathrm{B}_3$ with 
\begin{equation}{\scriptsize
\mathcal{I}_4=\begin{pmatrix} s_6/2& s_L& 0& 0& 0\\-s_R/2& 0& s_L/2& 0& 0\\0& -s_R/2& 0& s_L/2& 
  0\\0& 0& -s_R/2& 0& s_L/2\\0& 0& 0& -s_R& -s_6/2
 \end{pmatrix}\,,}
\end{equation}

\section{Anomaly cancelation in 6D models}\label{App:AnomalyCancel}

In Section~\ref{Sec:6DIIBCompactif} we computed the number of charged and neutral hypermultiplets. We distinguish the two cases with and without invariant brane:

\begin{itemize}
\item If there is the invariant brane, then
\begin{eqnarray}\label{totHwithInvD7}
 H &=& 14+\frac{\bar{K}}{2} + \left(\frac{a+b+1}{2}\right)^2[s_L]^2 - \frac12\left(\frac{a+b+1}{2}\right)[s_L]\cdot \bar{K} + \\
  && + ([x_1]+[y_1]+[w_2])\cdot\left( [x_1]+[y_1]+[w_2] +2 \left(\frac{a+b+1}{2}\right) [s_L]-\frac{\bar{K}}{2}  \right) \:.\nonumber
\end{eqnarray}
The tadpole cancellation condition 
\begin{equation}
[\mathcal{D}_{D7_1}]+[\mathcal{D}_{D7_1'}]+[\mathcal{D}_{D7_2}]+[\mathcal{D}_{D7_2'}]+  [\mathcal{D}_{D7_{\rm inv}}]  = 8[O7]
\end{equation}
constrain the classes $[x_1]$, $[y_1]$ and $[w_2]$:
\begin{equation}\label{xywD7tadp}
 [x_1] + [y_1] + [w_2] = 4\bar{K} - \left( \frac{a+b+1}{2} \right) [s_L] \:.
\end{equation}
Plugging \eqref{xywD7tadp} into \eqref{totHwithInvD7} one obtains $H=14+\frac{29}{2}\bar{K}^2$.
\item When there is no invariant brane we have
\begin{eqnarray}\label{totHwithoutInvD7}
 H &=& 14+\frac{\bar{K}}{2} + \left(\frac{a+b}{2}\right)^2[s_L]^2 - \frac12\left(\frac{a+b}{2}\right)[s_L]\cdot \bar{K} + \\
  && + ([x_1]+[y_1])\cdot\left( [x_1]+[y_1] +2 \left(\frac{a+b}{2}\right) [s_L]-\frac{\bar{K}}{2}  \right) \:. \nonumber
\end{eqnarray}
The D7-tadpole cancellation condition gives 
\begin{equation}\label{xywD7tadpNoinv}
 [x_1] + [y_1]  = 4\bar{K} - \left( \frac{a+b}{2} \right) [s_L]
\end{equation} 
and we obtain $H=14+\frac{29}{2}\bar{K}^2$.
\end{itemize}
In both cases this matches with the anomaly cancellation condition $H=V+273-29T$. Remember that in our setup $V=2$ where one vector multiplet is massless, while the other gets a mass by eating one (axionic) hypermultiplet. The total number $H$ that we computed includes such eaten hypermultiplet. If we count only massless hypermultiplets we should substract $1$ from both sides of the anomaly cancellation condition, and the match will be still valid.

\

Next we consider the mixed $\U1$ gravitational anomaly as well as the pure $\U1$ anomaly. 
The corresponding anomaly cancellation conditions read 
\beq\label{a1}
-\frac{1}{6}\sum_Q n_Q\, Q^2 =\frac{1}{2}\,\mathsf{a}\cdot\mathsf{b}\,,
\eeq
and 
\beq\label{a2}
\frac{1}{3}\sum_Q n_Q\, Q^4 =\frac12\,\mathsf{b}\cdot \mathsf{b}\,,
\eeq
with 
\beq
\mathsf{a}=\bar{K}\quad {\rm and }\quad \mathsf{b}=-\frac{a^2([D7_1]+[D7_1'])+b^2([D7_2]+[D7_2'])}{\kappa^2}\,,
\eeq
where $\mathsf{b}$ for example can be derived from the CS couplings $\int_{D7_i}C_4\wedge F_i\wedge F_i$.
These conditions are analogous to the ones we typically have in F-theory setups with a slight difference in a factor $1/2$ on the right hand side of Eqs. \eqref{a1} and \eqref{a2} due to the fact that we are computing intersections on the double cover manifold $X$ instead of the F-theory base $B$ (while on the left hand side we are summing only on the projected spectrum: for example we are not counting both states from $D7_1\cap D7_2$ and states from $D7_1'\cap D7_2'$). The divisor $\mathsf{b}$ is the analogous of the Nèron-Tate height pairing $-\pi(\sigma(S_1)\cdot \sigma(S_1))$ \cite{Park:2011ji}, with $\sigma(S_1)$ being the Shioda map of the section $S_1$ associated to the massless $\U1$ symmetry, i.e. the extra section in addition to zero section $S_0$.

Again we consider two different cases depending on whether the invariant brane is present or not. 
\begin{itemize}
\item If there is an invariant brane, one can actually verify that
\beq
-\frac{1}{6}\sum_Q n_Q\, Q^2 =\frac{1}{2}\bar{K}\cdot \left(-\frac{a b(a+b) [s_L] + 2 a^2 [x_1] + 2 b^2 [y_1]}{\kappa^2}\right)\,,
\eeq
and 
\beq
\frac{1}{3}\sum_Q n_Q\, Q^4 =\frac{1}{2} \left(-\frac{a b(a+b)[s_L]+ 2 a^2 [x_1] + 2 b^2 [y_1]}{\kappa^2}\right)^2\,,
\eeq
where we have used the expression for the massless $\U1$ charge provided in~\ref{masslessU1chargeGEN} and Table \ref{tab:2branes1inv}. We can then identify the IIB version of the Nèron-Tate height pairing as
\begin{align}
-\mathsf{b}=\frac{a^2([D7_1]+[D7_1'])+b^2([D7_2]+[D7_2'])}{\kappa^2}=\frac{a b(a+b) [s_L] + 2 a^2 [x_1] + 2 b^2 [y_1]}{\kappa^2}\,.
\end{align}
\item If there is no invariant brane in the setup, the anomaly expressions we get are the following
\beq
-\frac{1}{6}\sum_Q n_Q\, Q^2 =\frac{1}{2}\bar{K}\cdot \left(-\frac{8 b^2 \bar{K} +(a^2-b^2)(b[s_L]+2[x_1])}{\kappa^2}\right)\,,
\eeq
and 
\beq
\frac{1}{3}\sum_Q n_Q\, Q^4 =\frac{1}{2} \left(-\frac{8 b^2 \bar{K} +(a^2-b^2)(b[s_L]+2[x_1])}{\kappa^2}\right)^2\,.
\eeq
Hence, the type~IIB version of the Nèron-Tate height pairing is given by
\begin{align}
-\mathsf{b}=\frac{a^2([D7_1]+[D7_1'])+b^2([D7_2]+[D7_2'])}{\kappa^2}=\frac{8 b^2 \bar{K} +(a^2-b^2)(b[s_L]+2[x_1])}{\kappa^2}\,,
\end{align}
where we have used the D7 tadpole cancellation condition \ref{xywD7tadpNoinv}.
\end{itemize}

\section{Matching Charged Loci in Charge 3 (II) Model}
\label{app:Charge3II}

We now apply the limit to the F-theory matter loci of the charge 3 (II) model. Let us first compute the expressions for the brane and image brane loci in the  the  Calabi-Yau $X$. Using the replacements: $(s_L,s_6,s_R)=(s_5,s_6,s_7)$, $(p_1,p_2)=(s_9,s_8)$, $(r_1,r_2,r_3,r_4)=(s_4,s_3,s_2,s_1)$, the ideals can be obtained from Eqs. \eqref{eq:Dp} and \eqref{eq:Dr}. We obtain the following prime decomposition for the divisors of interest
\begin{align}
D7_1 & \equiv  \{2 s_7 s_8 - 2 \xi s_9 - s_6 s_9, 2 \xi s_8 - s_6 s_8 + 2 s_5 s_9\}\,, \\
D7_1^\prime & \equiv \{2 s_7 s_8 +2 \xi s_9 - s_6 s_9, 2 \xi s_8 + s_6 s_8 - 2 s_5 s_9\}\,.
\end{align}
and 
\begin{align}
D7_2^\prime \,\, \equiv \,\, &  \{2  \xi s_4 s_6^2  - s_4 s_6^3  - 2  \xi s_4 s_5 s_7 - 2  \xi s_3 s_6 s_7 +3 s_4 s_5 s_6 s_7 + s_3 s_6^2  s_7 + 2  \xi s_2 s_7^2 - 2 s_3 s_5 s_7^2 \nonumber \\
& - s_2 s_6 s_7^2  + 2 s_1 s_7^3,  2  \xi s_4 s_5 s_6 - s_4 s_5 s_6^2  - 2  \xi s_3 s_5 s_7 + 2 s_4 s_5^2  s_7 + s_3 s_5 s_6 s_7 \nonumber \\
&+2  \xi s_1 s_7^2  - 2 s_2 s_5 s_7^2 + s_1 s_6 s_7^2, 2  \xi s_4 s_5^2  - s_4 s_5^2 s_6 - 2  \xi s_2 s_5 s_7 +2 s_3 s_5^2  s_7\\
& + 2  \xi s_1 s_6 s_7 - s_2 s_5 s_6 s_7 + s_1 s_6^2  s_7 - 2 s_1 s_5 s_7^2, 2  \xi s_3 s_5^2  -2 s_4 s_5^3 - 2  \xi s_2 s_5 s_6 \nonumber \\
&+ s_3 s_5^2  s_6 + 2  \xi s_1 s_6^2  - s_2 s_5 s_6^2  + s_1 s_6^3  - 2  \xi s_1 s_5 s_7 + 2 s_2 s_5^2  s_7 - 3 s_1 s_5 s_6 s_7\}\,, \nonumber \\
& \nonumber \\
D7_2 \,\, \equiv \,\, & \{2  \xi s_4 s_6^2  + s_4 s_6^3  - 2  \xi s_4 s_5 s_7 - 2  \xi s_3 s_6 s_7 - 3 s_4 s_5 s_6 s_7 -s_3 s_6^2  s_7 + 2  \xi s_2 s_7^2  + 2 s_3 s_5 s_7^2 \nonumber \\ 
& + s_2 s_6 s_7^2  - 2 s_1 s_7^3, 2  \xi s_4 s_5 s_6 + s_4 s_5 s_6^2 - 2  \xi s_3 s_5 s_7 - 2 s_4 s_5^2  s_7 - s_3 s_5 s_6 s_7  \nonumber \\
&+ 2  \xi s_1 s_7^2 + 2 s_2 s_5 s_7^2- s_1 s_6 s_7^2, 2  \xi s_4 s_5^2  + s_4 s_5^2  s_6 - 2  \xi s_2 s_5 s_7 - 2 s_3 s_5^2  s_7 \\
&+ 2  \xi s_1 s_6 s_7 + s_2 s_5 s_6 s_7 - s_1 s_6^2  s_7 + 2 s_1 s_5 s_7^2, 2  \xi s_3 s_5^2  + 2 s_4 s_5^3  - 2  \xi s_2 s_5 s_6  \nonumber \\
& - s_3 s_5^2  s_6 + 2  \xi s_1 s_6^2  + s_2 s_5 s_6^2  - s_1 s_6^3  - 2  \xi s_1 s_5 s_7 - 2 s_2 s_5^2  s_7 + 3 s_1 s_5 s_6 s_7\}\,. \nonumber 
\end{align}
Remember that in type~IIB we have charge 3 states at $D7_1\cap D7_1' \setminus D7_1\cap O7$, charge 2 states at $D7_1\cap D7_2'$ and charge 1 states at $D7_1\cap D7_2$ and $D7_2\cap D7_2' \setminus D7_2\cap O7$.

Let us see how we obtain these loci by applying the Sen limit, that in this case is given by \eqref{eq:wcch3II}, i.e. $s_{1,2,3,4}$ scales as $\epsilon^1$.
\begin{itemize}
\item The charge 3 states live at the locus 
\beq\label{eq:idd1}
V(I_{(3)}):=\{s_8 = s_9 = 0\}\,.
\eeq
In the Sen limit, $s_{8,9}$ do not scale with $\epsilon$. In fact,  we obtain the same locus in type~IIB. Because of the identification $(s_9,s_8)=(p_1,p_2)$, one sees that this locus is exactly when the brane and its image intersect away from the orientifold locus (fixed locus of the involution), see \eqref{DD'AwayO7}.
Hence we find agreement  with what predicted in Table~\ref{tab:Ch3II_IIB}. 
\item The charge 2 locus in F-theory is given by 
\begin{equation}\label{EqApp:I2ch3}
V(I_{(2)}):=\{s_4 s_8^3- s_3 s_8^2 s_9 + s_2 s_8 s_9^2 -s_1 s_9^3
 =s_7 s_8^2  + s_5 s_9^2\!-\! s_6 s_8 s_9 =0\}\backslash\ V(I_{(3)}) \:.
\end{equation}
The two polynomials are homogeneous if $\epsilon$, hence the locus is not modified by sending $\epsilon\rightarrow 0$. Let us check that this gives the intersection $D7_1 \cap D7_2'$. 
Intersecting the locus \eqref{EqApp:I2ch3} with the Calabi-Yau $X$ makes it split into three codimension two ideals. One of them corresponds to \eqref{eq:idd1}, that should be removed,  while the other two are mapped to each other under the involution $\xi \mapsto-\xi$. One can show that these are exaclty what one obtains from the instersection $D7_1 \cap D7_2$ and $D7_1' \cap D7_2'$. The full expression reads\footnote{The primary decomposition for$D7_1 \cap D7_2'$ includes an additional codimension three piece $\{s_2=s_6=s_5=0\}$ that is absent as we are working in 6D compactifications}: 
\begin{align}
D7_1 \cap D7_2\,\,=\,\,& \{2s_7 s_8 - 2\xi  s_9 - s_6 s_9, 2\xi  s_8 - s_6 s_8 + 2s_5 s_9, s_4 s_8^3  - s_3 s_8^2 s_9+ s_2 s_8 s_9^2 \nonumber \\
&  - s_1 s_9^3 , 2s_4 s_6 s_8^2  - 2s_4 s_5 s_8 s_9 - 2s_3 s_6 s_8 s_9 +2\xi  s_2 s_9^2  + 2s_3 s_5 s_9^2  + s_2 s_6 s_9^2\nonumber \\
& - 2s_1 s_7 s_9 , 2s_4 s_5 s_8^2  - 2s_3 s_5 s_8 s_9 + 2\xi  s_1 s_9^2  + 2s_2 s_5 s_9^2  - s_1 s_6 s_9^2 , 2s_4 s_6^2 s_8 \nonumber \\
& - 2\xi  s_4 s_5 s_9 - 2\xi  s_3 s_6 s_9 - 3s_4 s_5 s_6 s_9 - s_3 s_6^2 s_9 + 2\xi  s_2 s_7 s_9 +2s_3 s_5 s_7 s_9 \nonumber \\
&+ s_2 s_6 s_7 s_9 - 2s_1 s_7^2 s_9, 2s_4 s_5 s_6 s_8 - 2\xi  s_3 s_5 s_9 - 2s_4 s_5^2 s_9 - s_3 s_5 s_6 s_9 \nonumber \\
& + 2\xi  s_1 s_7 s_9 + 2s_2 s_5 s_7 s_9 - s_1 s_6 s_7 s_9, 2s_4 s_5^2 s_8 - 2\xi  s_2 s_5 s_9 - 2s_3 s_5^2 s_9 \nonumber \\
& + 2\xi  s_1 s_6 s_9 + s_2 s_5 s_6 s_9 -s_1 s_6^2 s_9 + 2s_1 s_5 s_7 s_9, 2\xi  s_4 s_6^2  + s_4 s_6^3  - 2\xi  s_4 s_5 s_7 \nonumber \\
&- 2\xi  s_3 s_6 s_7 - 3s_4 s_5 s_6 s_7 - s_3 s_6^2 s_7 + 2\xi  s_2 s_7^2  + 2s_3 s_5 s_7^2  + s_2 s_6 s_7^2  - 2s_1 s_7^3 , \nonumber \\
& 2\xi  s_4 s_5 s_6 + s_4 s_5 s_6^2  - 2\xi  s_3 s_5 s_7 - 2s_4 s_5^2 s_7 - s_3 s_5 s_6 s_7 + 2\xi  s_1 s_7^2  + 2s_2 s_5 s_7^2 \nonumber \\
& - s_1 s_6 s_7^2 , 2\xi  s_4 s_5^2  + s_4 s_5^2 s_6 - 2\xi  s_2 s_5 s_7 -2s_3 s_5^2 s_7 + 2\xi  s_1 s_6 s_7 + s_2 s_5 s_6 s_7\nonumber \\
& - s_1 s_6^2 s_7 + 2s_1 s_5 s_7^2 , 2\xi  s_3 s_5^2  + 2s_4 s_5^3  - 2\xi  s_2 s_5 s_6 - s_3 s_5^2 s_6 + 2\xi  s_1 s_6^2 \nonumber \\ 
&+ s_2 s_5 s_6^2  - s_1 s_6^3  -2\xi  s_1 s_5 s_7 - 2s_2 s_5 s_7 + 3s_1 s_5 s_6 s_7\} \:.  \label{eq:ch2IIB}
\end{align}
\item In F-theory, the charge 1 locus is given by 
\beq 
V(I_{(1)}):=\{y_{1}=f z_1^4+3 x_1^2=0\}\backslash\ (V(I_{(2)})\cup V(I_{(3)}))
\eeq 
where $x_1$, $y_1$ and $z_1$ are the section coordinates in the Weierstrass form. These coordinates, written in terms of the sections $s_i$, can be found in equation (B8), Appendix B of \cite{Klevers:2014bqa}. Since their expressions are very long, we will not report them here.

One can first take the expression for $y_1$ and apply the weak coupling limit~\eqref{eq:wcch3II} and show that at leading order in $\epsilon$ $y_1$ takes the form
\begin{equation}
y_1=\epsilon \frac12 A z_1^2+\mathcal{O}(\epsilon^2)\,,
\end{equation}
with 
\begin{align}
\begin{split}
A\,=&\,s_4 s_6^2 - (s_4 s_5 + s_3 s_6) s_7 + s_2 s_7^2) s_8^3 - 
 3 (s_4 s_5 s_6 + s_7 (-s_3 s_5 + s_1 s_7)) s_8^2 s_9 \\
 &+ 
 3 (s_4 s_5^2 - s_2 s_5 s_7 + s_1 s_6 s_7) s_8 s_9^2 + (-s_3 s_5^2 + s_2 s_5 s_6 - 
    s_1 s_6^2 + s_1 s_5 s_7) s_9^3
    \end{split}\nonumber
\end{align}
and 
\begin{equation}
z_1= s_7 s_8^2  + s_5 s_9^2\!-\! s_6 s_8 s_9\,,
\end{equation}
similarly for $f z_1^4+3 x_1^2$,
\begin{equation}
f z_1^4+3 x_1^2=\epsilon \frac12 B z_1^3+\mathcal{O}(\epsilon^2)\,,
\end{equation}
with 
\begin{align}
\begin{split}
B=&s_7 (-s_3 s_6^2 + 2 s_3 s_5 s_7 + s_2 s_6 s_7 - 2 s_1 s_7^2) s_8^3 + 
 3 s_7 (s_3 s_5 s_6 - 2 s_2 s_5 s_7 + s_1 s_6 s_7) s_8^2 s_9\\
 & + 
 3 s_7 (-2 s_3 s_5^2 + s_2 s_5 s_6 - s_1 s_6^2 + 
    2 s_1 s_5 s_7) s_8 s_9^2 + (s_6 (s_3 s_5^2 + s_6 (-s_2 s_5 + s_1 s_6))\\
    & + 
    s_5 (2 s_2 s_5 - 3 s_1 s_6) s_7) s_9^3 + 
 s_4 (s_6 s_8 - 2 s_5 s_9) ((s_6^2 - 3 s_5 s_7) s_8^2 - s_5 s_6 s_8 s_9 + 
    s_5^2 s_9^2)\,.
    \end{split}\nonumber
\end{align}
Therefore, at weak coupling the locus $\{y_1=f z_1^4+3 x_1^2=0\}$ becomes $\{ Az_1^2=Bz_1^3=0\}$. Since we have to subtract $V(I_{(2)})\cup V(I_{(3)})$, then $z_1\neq 0$ and we obtain that the charge one locus is fully captured by $\{A=B=0\}$. Considering the primary decomposition of this locus in the double cover, one obtains three codimension two irreducible components: the first one is 
\begin{align}\label{eq:id1}
\begin{split}
D7_2 \cap D7_2'\setminus D7_2\cap O7 \,\,=\,\,&\{s_4 s_6^2  - s_4 s_5 s_7 - s_3 s_6 s_7 + s_2 s_7^2 ,
 s_4 s_5 s_6 - s_3 s_5 s_7 +s_1 s_7^2, \\
 & \,\, s_4 s_5^2  - s_2 s_5 s_7 + s_1 s_6 s_7, s_3 s_5^2  - s_2 s_5 s_6 + s_1 s_6^2  - s_1 s_5 s_7\},
 \end{split} 
\end{align}
that, as indicated, corresponds to the component of $D7_2 \cap D7_2'$ that is away from the orientifold locus $\xi=0$. 
The other two loci are orientifold image of each other; one of them is given by the following expression
\begin{align}\label{eq:id2}
\begin{split}
D7_1 \cap D7_2'\,\,=\,\,& \{2s_7 s_8 - 2\xi s_9 - s_6 s_9, 2\xi s_8 - s_6 s_8 + 2s_5 s_9, 2\xi s_4 s_6^2  -s_4 s_6 ^3 - 2\xi s_4 s_5 s_7\\
& - 2\xi s_3 s_6 s_7 + 3s_4 s_5 s_6 s_7 + s_3 s_6^2 s_7 + 2\xi s_2 s_7^2 - 2s_3 s_5 s_7^2  - s_2 s_6 s_7^2  + 2s_1 s_7^3 ,\\
& 2\xi s_4 s_5 s_6 - s_4 s_5 s_6^2  - 2\xi s_3 s_5 s_7 + +2s_4 s_5^2 s_7 + s_3 s_5 s_6 s_7 + 2\xi s_1 s_7^2  - 2s_2 s_5 s_7^2\\
&  + s_1 s_6 s_7^2 , 2\xi s_4 s_5^2 - s_4 s_5^2 s_6 - 2\xi s_2 s_5 s_7 + 2s_3 s_5^2 s_7 + 2\xi s_1 s_6 s_7 - s_2 s_5 s_6 s_7 \\
&+s_1 s_6^2 s_7 - 2s_1 s_5 s_7^2 , 2\xi s_3 s_5^2  - 2s_4 s_5^3  - 2\xi s_2 s_5 s_6 + s_3 s_5^2 s_6 +2\xi s_1 s_6^2 \\
& - s_2 s_5 s_6^2+ s_1 s_6^3  - 2\xi s_1 s_5 s_7 + 2s_2 s_5^2 s_7 - 3s_1 s_5 s_6 s_7\}.
\end{split}
\end{align}
We then found that the charge one locus of the F-theory threefold, splits into the two charge one loci that are expected in the corresponding type~IIB model.
\end{itemize}

\section{Matching  Charged Loci in Charge 4 Model}
\label{app:Charge4}

In this appendix we illustrate the matching of the codimension two loci for the charged matter in F-theory and type~IIB. Recalling the identifications $(s_L,s_6,s_R)=(s_3,s_6,s_8)$, $(p_1,p_2)=(b_1,a_1)$, $(w_1,w_2,w_3)=(s_2,s_5,s_1)$ and $(q_1,q_2,q_3)=(d_2,d_1,d_0)$, one can see that for the discriminant locus $\Delta_1$ (see~\eqref{delta_remain}) the splitting into brane/image brane is governed by \eqref{EqD7p1p2}. For the discriminant locus $\Delta_2$ is an invariant brane (see \eqref{invD7ord1}) and and similarly for $\Delta_3$ that splits in the Calabi-Yau $X$ accoding to~\eqref{EqD7q1q2}. 
\begin{itemize}
\item The charge four locus is given by
\begin{equation}\label{eq:c4l}
V(I_{(4)}):=\{a_1=b_1=0\}\:.
\end{equation}
Note that in this case $D7_1$ is described in terms of $(p_1,p_2)=(a_1,b_1)$. According to~\eqref{DD'AwayO7}, the brane and its image intersect away from the orientifold over the locus $\{a_1=b_1=0\}$, so that the charge 4 locus in type IIB corresponds to $D7_1 \cap D7_1'$, in accordance with table \ref{tab:Ch34IIB}.
\item The charge three locus in F-theory is described by the variety
\begin{equation}
V(I_{(3)}):=\{\alpha =\beta = 0\}\backslash\ V(I_{(4)})\,
\end{equation}
with $\alpha$ and $\beta$ given in~\eqref{eq.alphabeta}. We notice that after taking the scaling $a_1\rightarrow \epsilon a_1$ and $b_1\rightarrow \epsilon b_1$, the polynomials $\alpha$ and $\beta$ are homogeneus of degree two in $\epsilon$ and hence the locus does not suffer modifications at weak coupling.One can show that in the Calabi-Yau $X$ the ideal $I_{(3)}$ decomposes into three prime ideals: The first one is given by \eqref{eq:c4l} and it should be removed. The remaining two are image to each other under $\xi\mapsto -\xi$. One of them is equal to the intersection of $D7_1$ and $D7_2'$ up to codimension three loci (the other is its orientifold image):
\begin{align}
D7_1\cap D7_2'= &\, \{2\xi b_1 - b_1 s_6 + 2a_1 s_8, 2\xi a_1 - 2b_1 s_3 + a_1 s_6, 2\xi d_2 s_6 - d_2 s_6^2- 2\xi d_1 s_8\nonumber \\
 &+ 2d_2 s_3 s_8 + d_1 s_6 s_8 - 2d_0 s_8^2, b_1 d_2 s_3 - a_1 d_2 s_6 - b_1 d_0 s_8 + a_1 d_1 s_8,\nonumber \\
 & 2\xi d_2 s_3 - d_2 s_3 s_6 - 2\xi d_0 s_8 + 2d_1 s_3 s_8 -d_0 s_6 s_8, b_1 d_1 s_3 - a_1 d_2 s_3\nonumber \\
 & - b_1 d_0 s_6 + a_1 d_0 s_8, 2\xi d_1 s_3^2 - 2d_2 s_3  - 2\xi d_0 s_6 + d_1 s_3 s_6 - d_0 s_6^2 \nonumber \\
 & + 2d_0 s_3 s_8, b_1 d_0^2 - a_1 b_1 d_1 + a_1^2 d_2\}\:.
\end{align}
\item The charge two locus is given by the following expression 
\begin{equation}
V(I_{(2)}):=
 \{t=\hat{z}=0\}\backslash\ V(I_{(4)})\cup V(I_{(3)})
\end{equation}
where $t$ and $\hat z$ are given in equations \eqref{eq:tt} and \eqref{eq:zz} respectively. In this case, the scalings with $\epsilon$ change the locus. Note that upon scaling $t$ and $\hat z$ take the following form 
\begin{equation}
t=\epsilon^4 \beta A+\mathcal{O}(\epsilon^5)\,, \quad \hat z= \epsilon^4 \beta B+\mathcal{O}(\epsilon^5)\,,
\end{equation}
with 
\begin{equation}
A=2 a_1 b_1 (d_1 s_3 - d_0 s_6) s_8 + a_1^2 s_8 (-d_2 s_3 + d_0 s_8) + 
 b_1^2 (d_2 s_3^2 - d_1 s_3 s_6 + d_0 s_6^2 - d_0 s_3 s_8)\,,
\end{equation}
and 
\begin{equation}
B=b_1^2 (d_1 s_3 - d_0 s_6) + 2 a_1 b_1 (-d_2 s_3 + d_0 s_8) + a_1^2 (d_2 s_6 - d_1 s_8)\,,
\end{equation}
therefore at weak coupling, the charge 2 ideal becomes $\{\beta A,\beta B\}$. This ideal decomposes into three codimension 2 ideals when intersected with the type IIB CY $X$. The first one corresponds to the intersection of $D7_2$ and its image away from the orientifold locus:
\begin{align}
D7_2\cap D7_2'\backslash\, D7_2 \cap O7= &\, \{d_2 s_6 - d_1s_8, d_2s_3 - d_0s_8, d_1 s_3 - d_0s_6\}\:,
\end{align}
as expected for brane/image-brane wrapping a divisor of order 2 in $s_{6,L,R}$, see \eqref{order2D7D7pintersection}.
The remaining two ideals are mapped to each other under the orientifold involution. One can show that one of these ideals corresponds to 
\begin{align}
D7_1\cap D7_{\rm inv}= &\, \{2\xi  b_1 -  b_1  s_6 + 2a_1  s_8, 2\xi a_1 - 2 b_1  s_3 + a_1  s_6, 2\xi  d_2  s_6 -  d_2  s_6^2\nonumber \\
&  - 2\xi  d_1  s_8 + 2 d_2  s_3  s_8 +  d_1  s_6  s_8 - 2 d_0  s_8^2 ,  b_1  d_2  s_3 - a_1  d_2  s_6 -  b_1  d_0  s_8 \nonumber \\
&+ a_1  d_1  s_8, 2\xi  d_2  s_3 -  d_2  s_3  s_6 - 2\xi  d_0  s_8+ 2 d_1  s_3  s_8 -  d_0  s_6  s_8,  b_1  d_1  s_3\nonumber  \\
&- a_1  d_2  s_3 -  b_1  d_0  s_6 + a_1  d_0  s_8,2\xi  d_1  s_3 - 2 d_2  s_3^2  - 2\xi  d_0  s_6 +  d_1  s_3  s_6\nonumber \\
& -  d_0  s_6^2  + 2 d_0  s_3  s_8,  b_1^2  d_0 -a_1  b_1  d_1 + a_1^2  d_2\}\,.
\end{align}
Hence, the splitting of the F-theory matter locus coincides with the expectations from the type IIB side (see Table \ref{tab:Ch34IIB}). 
\item The charge one locus is written as
\begin{equation}
V(I_{(1)}):=\{\hat y=f \hat z^4+3 \hat x^2=0\}\backslash\ V(I_{(4)})\cup V(I_{(3)})\cup (V(I_{(2)})
\end{equation}
in terms of the coordinates $[\hat x:\hat y:\hat z]$ for the section in the Weierstrass polynomial. The expressions for these, in terms of the base sections $a_1$, $b_1$ $s_i$, $d_i$, are very long and can be found in a Mathematica notebook (Charge4Model.nb) as part of the ancillary files of the arXiv post\footnote{\url{https://arxiv.org/src/1711.03210v1/anc}} of ref. \cite{Raghuram:2017qut}. 
Following an analogous procedure as the one outlined in this and in the previous appendix one can apply the weak coupling limit to this locus and find that it splits in type IIB to the intersections $D7_2\cap D7_2$ and $D7_2\cap D7_{\rm inv}$. 
\end{itemize}


\bibliographystyle{utphys}	
\providecommand{\href}[2]{#2}\begingroup\raggedright\endgroup

\end{document}